\documentclass[pra,a4paper,nofootinbib,twocolumn,showpacs,preprintnumbers,amsmath,amssymb,floatfix,amstex,superscriptaddress]{revtex4}
\newcommand{\ket}[1]{|#1\rangle}                	%
\newcommand{\bra}[1]{\langle #1}               		%
\usepackage{graphicx,graphics,wrapfig,rotating}     	
\usepackage{dcolumn}                    		
\usepackage{bm,fancybox}                		
\usepackage{times,euscript,eufrak,oldgerm}              %
\usepackage[german,english]{babel}			%
\usepackage{psfrag}
\begin{document}
\title{Decoherence models and their effects on quantum maps and algorithms} %
\author{Mario Leandro Aolita}
\altaffiliation[Present address: ]{%
Instituto de F\'\i sica, Universidade Federal do Rio de Janeiro. Caixa Postal 
68528, 21941-972 Rio de Janeiro, RJ, Brasil.\\
Email address: \url{aolita@if.ufrj.br}}
\affiliation{%
Departamento de F\'{\i}sica, Comisi\'{o}n Nacional de Energ\'{\i}a At\'{o}mica.
Avenida del Libertador 8250 (C1429BNP), Buenos Aires, Argentina.
}
\author{Marcos Saraceno}
\email{saraceno@tandar.cnea.gov.ar}
\affiliation{%
Departamento de F\'{\i}sica, Comisi\'{o}n Nacional de Energ\'{\i}a At\'{o}mica.
Avenida del Libertador 8250 (C1429BNP), Buenos Aires, Argentina.
}%
\affiliation{%
Escuela de Ciencia y Tecnolog\'\i a, Universidad Nacional de San
Mart\'\i n. Alem 3901 (B1653HIM), Villa Ballester, Provincia de  Buenos Aires,
Argentina.}
\date{\today}%
\begin{abstract}
In this work we study several models of decoherence and how different quantum maps and algorithms react when perturbed
by them. Following closely Ref.\ \cite{Yo}, generalizations of the three paradigmatic one single qubit quantum channels
 (these are the depolarizing channel, the phase damping channel and the amplitude damping channel) for the case of
an arbitrarily-sized finite-dimensional Hilbert space are presented, as well as other types of noise in phase
space. More specifically, Grover's search algorithm's response to decoherence is analyzed; together with
those of a family of quantum versions of chaotic and regular classical maps (the baker's map and the cat maps). A
relationship between how sensitive to decoherence a quantum map is and the degree of complexity in the dynamics
of its associated classical counterpart is observed; resulting in a clear tendency to react the more decoherently
the more complex the associated classical dynamics is.    

\end{abstract}
\maketitle

\section{Introduction}
The fundamental dualism inherent to quantum mechanics' axiomatic base, (deterministic) unitary evolution
and (probabilistic) measurement, has been disturbing physicists (among others) ever since the very conception of quantum
mechanics itself \cite{epr}. This very fundamental problem is in turn related to a potential practical application which has become a major
subject of research in the last two decades: quantum information processing, for which the maintenance of the
relevant system's coherence is a necessary condition. Therefore, understanding the quantum-to-classical
transition and, in particular, decoherence (loss of quantum information) has become of central interest \cite{zurek1}.
\par Decoherence's basic idea \cite{preskill,chuang} consists of assuming that a closed system's evolution is always
unitary and that the system of interest (e.g. a quantum computer), which from now on will be called the principal system, is never
isolated from the environment.     
In this scheme the composite system formed by the principal system and its surrounding environment (namely, the rest of the
universe) is a closed system and thus evolves unitarily. But if the
accessible information is only that corresponding to  the principal system, or if the access to
the environment's information is just an averaged one, by tracing out the environment's degrees of freedom one obtains for the reduced density
 operator of the principal system, $\hat{\rho}$, the following evolution
\begin{equation}
	\label{eq:supop}
    \hat{\rho}'\equiv \$(\hat{\rho}) \equiv \sum_{\mu}\hat{M}_{\mu}\hat{\rho}\hat{M}_{\mu}^{\dagger} \ \ ,
\end{equation}
where $\hat{\rho}'$ is the transformed reduced density operator \cite{comentario} of the principal system after a certain time $t$. The right hand side of equation (\ref{eq:supop}) is the so called
Kraus representation (or operator sum representation) of a superoperator $\$$,
where $\hat{M}_{\mu}$ are arbitrary operators satisfying the trace preserving condition
 $\sum_{\mu}\hat{M}_{\mu}^{\dagger}\hat{M}_{\mu}=\hat{I}$. Equation (\ref{eq:supop}) also guarantees that $\$$ be a completely positive map and that  the
 evolution be Markovian \cite{Kraus}. A superoperator is thus a completely positive trace preserving linear map (CPTPLM), and  the most
 general markovian evolution is ruled  by such a map (the unitary evolution is just a particular case, when there is only one Kraus operator in the sum).
 During the rest of the paper we will
stick to the following notation for $\$$:
\begin{equation}
   \$\equiv\sum_{\mu}\hat{M}_{\mu}\odot\hat{M}_{\mu}^{\dagger} \ \ ,
\end{equation} 
where the symbol symbol ``$\odot$'' is defined by comparison with (\ref{eq:supop}).
\par In  Ref. \cite{Yo} generalizations of the depolarizing and phase damping channels to the case of
an arbitrarily-sized finite-dimensional Hilbert space were introduced. They were studied in the
context of the chord representation \cite{chordrep} and in a (discrete) phase-space-based approach. 
In particular it was found that, using the symplectic invariance of the chord operators, special decoherence models could be constructed that produced decoherence
towards selected ``pointer states'' by diffusing on phase space lines.
\par Here we will present a brief
revision of these  noise models and include the generalization to the case of  amplitude damping channels. The focus of the present paper, 
in contrast to what was done in \cite{Yo}, will be to study the effect of these noise models on the evolution of otherwise unitary maps. We study the spectra of the 
superoperators and use DPS representations to display the noisy evolution.  
\par More specifically, we study various quantized maps that have different classical limits with the purpose of understanding how their underlying regularity or hyperbolicity react to the noise. We also  
 study the Grover's search algorithm, which has no classical
analog but is one of the simplest examples of quantum algorithms that can outperform their classical analogs.
\par The paper is structured as follows. In Sec. \ref{GDC} and \ref{GPDC} we review the generalized depolarizing and phase damping channels, show their actions in the DPS and
introduce quantum circuits to implement them. Sec. \ref{GADC} is devoted to the generalization of the amplitude damping channel. After that, in Sec. \ref{spectra}, a reinterpretation of these channels' actions is provided in terms of their spectra. The
response of unitary maps when decoherence is introduced is studied in Sec. \ref{Noise}. In particular, in Sec. \ref{NoisyGrover} we study the action in DPS and the
spectra of these noisy channels composed with the unitary Grover transformation; and in Sec. \ref{NoisyMaps} the composition is made with the quantum baker's map and
with the quantum cat maps. For these cases, the loss of purity of quantum states is quantified by means of the linear entropy. Finally, we conclude with our results' 
implications in Sec. \ref{Conclu}.
\section{Generalized noisy channels}
		\label{sec:models}
\par Quantum information processing's non trivial advantages over the classical arise when considering a large Hilbert space. Thus in order to study
the noisy evolution of a non trivial quantum information processor, first thing we need to do is develop  generalizations of the one-single-qubit 
quantum channels already mentioned to the case of an arbitrarily-sized finite-dimensional Hilbert space.
\subsection{Generalized depolarizing channel.}
		\label{GDC}
\par In \cite{Yo} such a generalization was proposed for the depolarizing channel.  The
Kraus representation of the superoperator obtained therein in the chord representation is:
\[
   \$_{DC}\equiv(1-\epsilon)\hat{T}_{0}\odot\hat{T}^{\dagger}_{0}+\frac{\epsilon}{N^{2}}\sum_{\alpha=0}^{N^{2}-1}\hat{T}_{\alpha}\odot\hat{T}^{\dagger}_{\alpha}
\]
\begin{equation}
   \equiv(1-\epsilon)\hat{I}\odot\hat{I}+\epsilon\tilde{\$}_{DC},
\end{equation}
where the subindex $\alpha$ is a shorthand notation for one of the $N^{2}$ DPS points $(q,p)$.  The map $\tilde{\$}_{DC}$ is also a properly normalized superoperator and its physical interpretation is
quite simple from a phase space point of view. It performs all possible displacements with equal weight thus averaging over all DPS points. It can therefore be
considered within the family of the ``diffusive superoperators'' \cite{garma,garma2,nonn} but with a very particular type of diffusion, one in which the
density matrix is spread over the whole DPS uniformly. So, with probability $1-\epsilon$,  $\$_{DC}$ leaves the state unchanged while, with probability $\epsilon$, it
averages it  over all DPS points. 

\begin{figure}[h!]
  \begin{center}
    \scalebox{0.5}{%
    \includegraphics*[59,88][553,640]{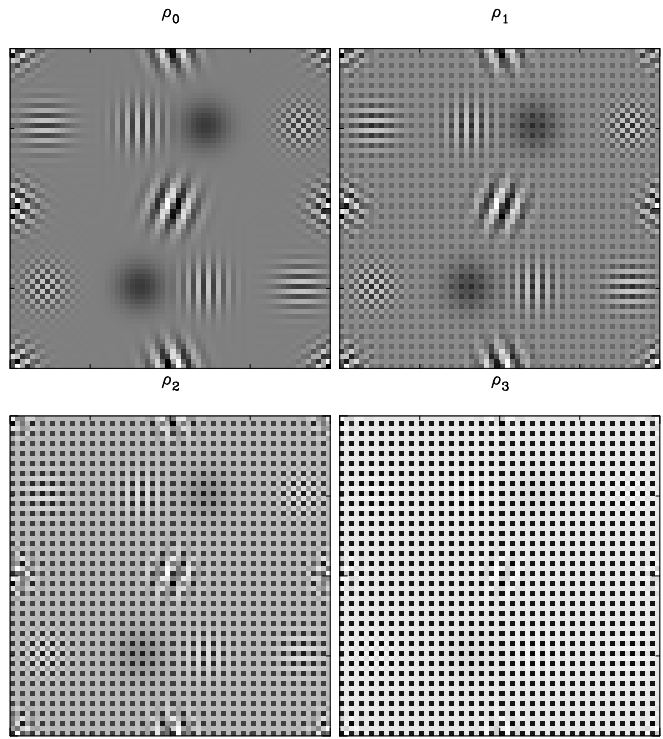}}
  \end{center}
\caption{\footnotesize Wigner function representation of the evolution of an initial superposition of two coherent states under the action  of $\$_{DC}$. The vertical and horizontal
axes correspond to the coordinates $p$ and $q$ respectively, and the gray scale intensity is proportional to the value of the Wigner function in that 
point, being white the most negative value and black the most positive one. In the first graphic both the interference fringes due to the periodicity
of the DPS and the quantum interference fringes (the ones present only in coherent superpositions) can be seen; after
applications of $\$_{DC}$ these interferences disappear and, in the final graphic, they are completely replaced by the chessboard-like Wigner function 
corresponding to the completely mixed state $\frac{\hat{I}}{N}$, the microcanonical distribution.\label{fig:F32abajo}}
\end{figure} 
\par In Fig. \ref{fig:F32abajo} the action of $\$_{DC}$ on a quantum state can be seen in phase space. There we have plotted the discrete Wigner function of
the initial ``cat state'' $\ket{\psi}=\frac{1}{\sqrt{2}}(\ket{(q,p)=(0.4,0.25)}+\ket{(q,p)=(0.6,0.75)})$, where $\ket{(q,p)}$ represents the coherent
state centered at the phase space point $(q,p)$, after no application of the map, $\hat{\rho}_{0}$, and after one, two and three applications,
$\hat{\rho}_{1}$, $\hat{\rho}_{2}$ and $\hat{\rho}_{3}$, respectively. The size of the Hilbert space is $N=32$ (five qubits) and the decoherence
parameter used is $\epsilon=0.8$. In the first plot we can observe the two gaussian black spots at the centers of the two coherent states, 
the quantum interferences between the two coherent states (like that observed right in between the two black spots) and the interferences
with the images that come from the periodic boundary conditions chosen in  the discretization of the phase space (which can now be thought of as a
torus). After three applications of
the map all interferences have disappeared and the resulting graphic is the chessboard-like Wigner function that corresponds to the state $\frac{\hat{I}}{N}$, the
 microcanonical distribution, the minimal (classical) information situation. So not only does $\$_{DC}$ provoke loss quantum information (decoherence) but
also of the classical one. The same result is observed when applying the map to position or momentum  states superpositions. Thus we see
that the generalized depolarizing channel has the completely mixed state as its only invariant sate (as is also the case for the one-single-qubit channel). 
\par
\begin{figure}[h!]
  \begin{center}
    \scalebox{1.2}{%
    \includegraphics*[0,0][197,86]{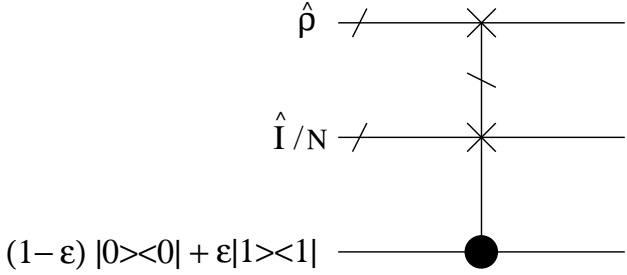}}
  \end{center}
\caption{\footnotesize Circuit implementation of the generalized depolarizing channel for a principal system consisting of a set of $n$ qubits
(upper line). The environment is modeled with another set of $n$ qubits initialized in the completely mixed state (central line). There is also an ancilla qubit (lower
line) that works as the control of the $n$-qubit-controlled-SWAP gate.\label{fig:F38}}
\end{figure}

 In view of the latter, it is now easy for us to think of a circuit that implements the action of this superoperator for the case of $N=2^{n}$ (that is, for a set
of $n$ qubits). All we need is a circuit that leaves the density matrix of the $n$ qubits unchanged with the correct probability ($1-\epsilon$), or
that replaces it by the state $\frac{\hat{I}}{N}$ (with probability $\epsilon$) \cite{comentario2}. And that is precisely
what the circuit in Fig. \ref{fig:F38} does. There we can observe an upper line marked with the symbol ``/'', that means transport of $n$ qubits (the principal
system), a central line that also transports $n$ qubits (which represent the environment, initialized in the state $\frac{\hat{I}}{N}$) and a lower line that
transports just one ancilla qubit (initialized in the state $(1-\epsilon)\ket{0}\bra{0}|+\epsilon\ket{1}\bra{1}|$). This ancilla qubit
works as the control for the $n$-qubit-controlled-SWAP gate, that does nothing when the state of the control is $\ket{0}$ (probability $1-\epsilon$) or exchanges the
states of the two $n$-qubit-systems when the control is in state $\ket{1}$ (probability $\epsilon$).
\subsection{Generalized phase damping channel and phase damping channel on a line.}
   \label{GPDC}
The generalizations we present here were also developed in \cite{Yo}. Written in terms of the skew projectors (or transition projectors) $\hat{P}_{ij}\equiv\ket{i}\bra{j}|$ 
(being $\ket{i}$ the $i^{th}$-member of the computational basis, $0\le i\le N-1=2^{n}-1$), the  Kraus representation of the generalized phase damping channel is
\begin{equation}
     \label{eq:PDC}
  \$_{PDC}=(1-\epsilon)\hat{I}\odot\hat{I}+\epsilon\sum_{i,j=1}^{N}
  C_{ij}\hat{P}_{ij}\odot\hat{P}_{ij}^{\dagger},
\end{equation}
where $C_{ij}$ must be an $N \times N$ real symmetric positive semidefinite stochastic matrix. If we choose $C_{ij}=\delta_{ij}$, where $\delta_{ij}$ is the
Kronecker delta, we obtain a superoperator only with diagonal projectors $\hat{P}_{ii}$(from now on we will refer to generalized damping channel
as the generalization for this choice of $C_{ij}$, other choices will be explicitly stated). In this case the
action of the map is to do nothing to the diagonal elements of the density matrix it acts on and to multiply its non diagonal elements by the factor
$1-\epsilon$, exactly as the one-single-qubit phase damping channel does \cite{Yo,preskill,chuang}. For more general choices of $C_{ij}$ the action on the non diagonal elements is
 the same, but the diagonal ones are then altered.
\par This superoperator is also diagonal in the chord representation, it is:
\[
   \$_{PDC}\equiv(1-\epsilon)\hat{T}_{0}\odot\hat{T}^{\dagger}_{0}+\frac{\epsilon}{N}\sum_{p=0}^{N-1}\hat{T}_{(0,p)}\odot\hat{T}^{\dagger}_{(0,p)}
\]
\begin{equation}
   \equiv(1-\epsilon)\hat{I}\odot\hat{I}+\epsilon\tilde{\$}_{PDC}.
\end{equation}
Again, $\tilde{\$}_{PDC}$ is a properly normalized diffusive-superoperator-like map, but instead of performing a uniform diffusion over the whole DPS as
$\tilde{\$}_{DC}$ does, it displaces the density matrix to all the $N$ points along the line of equation $q=0$ with the same weight $\frac{1}{N}$. It diffuses the
density matrix along the vertical direction. So $\$_{PDC}$ does nothing with probability $1-\epsilon$
and  projects the Wigner function onto the vertical lines that correspond to the position states (with which we have arbitrarily identified the computational states) 
with probability $\epsilon$. But the vertical direction is not at all a preferential one, it only appears because of the arbitrary identification we have made of the
computational states with the position states. The general case was obtained in \cite{Yo} with what was called the generalized phase damping channel on a line
$L_{n_{1},n_{2},n_{3}}$ \cite{comentario3}:
\[
   \$_{PDC_{(n_{1},n_{2},n_{3})}}\equiv(1-\epsilon)\hat{T}_{0}\odot\hat{T}^{\dagger}_{0}+\frac{\epsilon}{R}\sum_{L_{n_{1},n_{2},n_{3}}}\hat{T}_{(q,p)}\odot\hat{T}^{\dagger}_{(q,p)}
\]
\begin{equation}
   \equiv(1-\epsilon)\hat{I}\odot\hat{I}+\epsilon\tilde{\$}_{PDC_{(n_{1},n_{2},n_{3})}};
\end{equation}
where $R$ is the number of points $(q,p)$ in $L_{n_{1},n_{2},n_{3}}$ (not necessarily $N$ \cite{Miquel}) and the sum is performed over them.
\par
\begin{figure}[h!]
\begin{center}
    \scalebox{0.5}{%
    \includegraphics*[59,88][553,640]{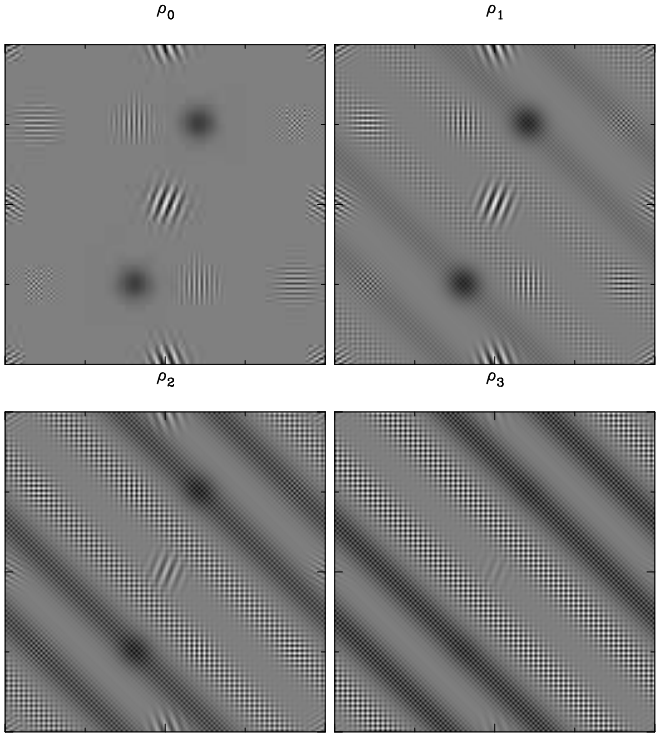}}
\end{center}
\caption{\footnotesize Evolution of the usual initial cat state upon application of the map $\$_{PDC_{(1,-1,0)}}$. The value of the decoherence parameter $\epsilon$ is
the usual too, but now the dimension of the Hilbert space has been taken as $N=64$ (six qubits) for a better appreciation of this map's action. In the last panel we
can see how the diffusion along the line $p=-q$ has almost made the quantum interferences disappear, the remaining state is an incoherent superposition of the
projections of the initial state on the eigenstates of  $\hat{T}_{(-1,1)}$, which are this superoperator's pointer states. 
\label{fig:F42arriba}} 
\end{figure}

 This superoperator's action in phase space can be appreciated in Fig.
\ref{fig:F42arriba}, where we have plotted the Wigner function representation of the evolution of an initial cat state for the case of 
$n_{1}=1$, $n_{2}=-1$ and $n_{3}=0$.
As much as the simple $\$_{PDC}$ ($=\$_{PDC_{(0,n_{2},0)}}$) does with the position states (which are the
eigenstates of $\hat{T}_{(0,p)},p\in\Bbb{Z}$), $\$_{PDC_{(1,-1,0)}}$
does nothing to the state with probability $1-\epsilon$ or takes it to an incoherent superposition (note how at the final plot the central quantum interference fringes have been
practically removed) of its projections onto the eigenstates of $\hat{T}_{(1,-1)}$, this
map's pointer states. In a more general direction of diffusion,   $\$_{PDC_{(n_{1},n_{2},0)}}$ acts exactly as $\$_{PDC}$ but with the computational basis being that of the
eigenstates of $\hat{T}_{(n_{1},n_{2})}$ instead of the position basis; that is, it does nothing to the diagonal elements of a density matrix
written in the basis of $\hat{T}_{(n_{1},n_{2})}$'s eigenstates and multiply its non diagonal ones by the factor $1-\epsilon$.
 Thus the pointer states for this model can be selected by choosing the diffusion line. In the most general case when the ordinate at the origin $n_{3}$ is different from zero, the
action $\$_{PDC_{(n_{1},n_{2},n_{3})}}$ becomes a combination of the one just described for $\$_{PDC_{(n_{1},n_{2},0)}}$ plus a unitary translation \cite{Yo,tesis}.
\par
\begin{figure}[h!]
  \begin{center}
    \scalebox{0.6}{%
    \includegraphics*[0,0][430,260]{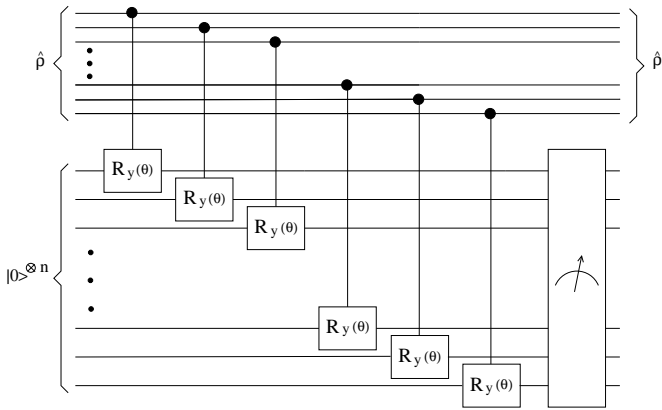}}
  \end{center}
\caption{\footnotesize Circuit model implementing the generalized phase damping channel on a system of $n$ qubits. If it is  $\cos{\frac{\Theta}{2}}=1-\epsilon$ then the
non diagonal elements of the principal system's transformed density matrix $\hat{\rho}'$ are multiplied by $1-\epsilon$; while the diagonal ones remain the same.\label{fig:F39}} 
\end{figure}

 The circuit shown in Fig. \ref{fig:F39} implements the generalized phase damping channels (if the choice for the computational basis is the position basis the resulting
superoperator is $\$_{PDC}$, if the chosen computational basis is that of the eigenstates of $\hat{T}_{(n_{1},n_{2})}$ then the resulting superoperator is $\$_{PDC_{(n_{1},n_{2},0)}}$).
 It is composed by $n$ controlled-rotation gates $R_{y}(\theta)$
 (that implement the unitary rotation $\hat{R}_{y}(\theta)$ 
around the $y$-axes in an angle $\theta$ if and only if the state of the control is $\ket{1}$) acting independently on every one of the $n$
pairs of qubits one from the principal system (upper lines) and one from the environment (lower lines). The environment is initialized in the state $\ket{0}^{\otimes n}$
(which is a short-hand notation for the composite state in which all $n$ qubits are in state $\ket{0}$). The last gate is a projective measurement in the product basis
$\{\ket{0}\otimes...\ \ ...\ket{0}\otimes\ket{0},\ket{0}\otimes...\ \ ...\ket{0}\otimes\ket{1},...\ \ ...,\ket{1}\otimes...\ \ ...\ket{1}\otimes\ket{1}\}$, which is
nothing  but $n$ independent one qubit projective measurements. It can be easily shown \cite{tesis} that the effect of this circuit on a density matrix $\hat{\rho}$ written in the
computational basis is to leave the diagonal elements unchanged and to multiply the  non diagonal ones by the factor $\cos{\frac{\theta}{2}}$. So, by choosing  $\theta$ such that 
$\cos{\frac{\theta}{2}}=1-\epsilon$ we get the desired
action for the circuit implementation of the generalized phase damping channel.
\subsection{Generalized amplitude damping channel}
   \label{GADC}
\par We now turn to a generalization of the one qubit amplitude damping channel. 
This is a schematic model for the process of decay of a two level excited atom by means of spontaneous emission of a photon
\cite{preskill,chuang}. Its Kraus form is:
\[
  \$^{1}_{ADC}=(\ket{0}\bra{0}|+\sqrt{1-\epsilon}\ket{1}\bra{1}|)\odot(\ket{0}\bra{0}|+\sqrt{1-\epsilon}\ket{1}\bra{1}|)
\]
\begin{equation}
  + \ \ \epsilon\ket{0}\bra{1}|\odot\ket{1}\bra{0}|\ \ .
\end{equation}
Again, $\ket{0}$ and $\ket{1}$ are the computational states which, in this case, are the ground and excited states of the atom, respectively. The density matrix' elements transform according to
\begin{equation}
  \left(\begin{array}{cc}
		\rho_{00} & \rho_{01} \\
		\rho_{10} & \rho_{11}
	 \end{array}
  \right) 
  \rightarrow
  \left(\begin{array}{cc}
		\rho_{00}+\epsilon\rho_{11} & \sqrt{(1-\epsilon)}\rho_{01} \\
		\sqrt{(1-\epsilon)}\rho_{10} & (1-\epsilon)\rho_{11}
	 \end{array}
  \right) \ \ . 
    \label{eq:adc}
\end{equation} 
The non-diagonal elements are damped but (in contrast to what happened with $\$_{PDC}$)
here the amplitude of the excited state is also damped and the atom eventually ends up in the ground state.
\par We now allow the atom to
have $N$ energy levels (which will be used as the computational states) and the field 
at  non-zero-temperature in such a way that there are photons in the
 field and the atom can absorb one of them
making a transition to an upper energy level. In the one-single-qubit case the transition from the excited to the ground state is carried out by an annihilation
operator accompanied by a factor $\epsilon$ (Kraus operator $\sqrt{\epsilon}\ket{0}\bra{1}|$). Now we will need Kraus operators that take any computational state $\ket{i}$, 
with $i=0,1,2...   ...N-1$, to the $\ket{i+\mu}$ with a certain transition probability $p_{i , i+\mu}$. It is immediate to convince one self that the following Kraus
operators will do:
\begin{equation}
  \hat{M}_{\mu}=\sum_{i=0}^{N-1}\sqrt{p_{i , i+\mu}}\hat{P}_{i+\mu , i} \ \ .
    \label{eq:adcKruasop}
\end{equation}
The transition probability matrix $p_{i, i+\mu}$ must be real, positive semidefinite, with all its elements lower or equal to $1$ and greater or equal to  $0$ (it
must be a probability), stochastic, and such that $p_{i , i+\mu}\equiv 0$ for all $0>i+\mu$ or $ i+\mu>N-1$ (so that there is a null probability for any  state to be
excited up beyond the $\ket{N-1}$ or to decay down beyond the $\ket{0}$). It is straightforward to show \cite{tesis} that if the matrix $p_{i , j}$ is also chosen to be
symmetric, the resulting superoperator is self adjoint. In particular we chose this matrix to be Gaussian with a mean width $\epsilon$: 
$p_{i , j}=p_{i , j}(|i-j|)\propto e^{\frac{(i-j)^{2}}{2\epsilon^{2}}}$, so that the closer two states are in energy the greater the probability of transition between one
another is. Thus the generalized amplitude damping channel will have Kraus operators that raise any
state a maximum of $N-1$ levels ($\mu>0$), others that will lower it a minimum of $N-1$ levels ($\mu<0$) and another that leaves the state unchanged ($\mu=0$). Its
Kraus representation is the following,
\begin{equation}
  \$_{ADC}=\sum_{\mu=-N+1}^{N-1}\hat{M}_{\mu}\odot\hat{M}_{\mu}^{\dagger} \ \ ,
\end{equation}
with $ \hat{M}_{\mu} $ given by (\ref{eq:adcKruasop}).
\begin{figure}[h!]
\begin{center}
    \scalebox{0.5}{%
    \includegraphics*[59,88][553,640]{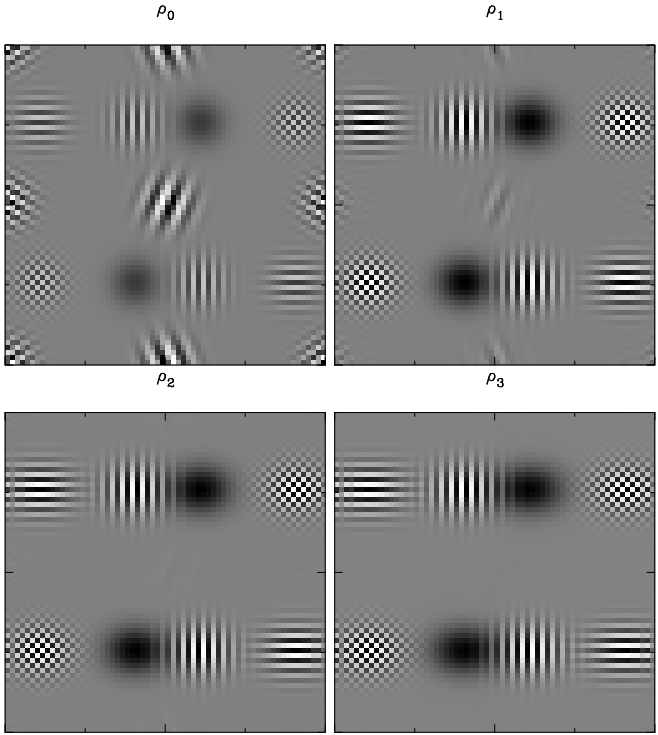}}
\end{center}
\caption{\footnotesize Evolution in phase space of the usual initial state upon application of  $\$_{ADC}$. The central quantum interference fringes rapidly disappear
and all other spots are stretched along the horizontal direction.
\label{fig:F35medio}} 
\end{figure} 
\par In Fig. \ref{fig:F35medio} we can see the phase space evolution of the usual cat state under the action of $\$_{ADC}$ for the usual parameters choice. We see there
that already in the first iteration the quantum interferences have almost disappeared, showing how strong a  decoherer this superoperator is. We can also see in the last
panel how the surviving spots have been stretched a little bit along the horizontal direction. The latter can be understood as follows, $\$_{ADC}$ picks any computational state  and 
displaces it symmetrically to all its neighbours with a certain probability, but as we have arbitrarily chosen the computational states to be the position states this
shows in the plot  as a spreading of the initial spots in the horizontal direction.   
\par
\begin{figure}[h!]
  \begin{center}
    \scalebox{1.2}{%
    \includegraphics*[0,0][174,66]{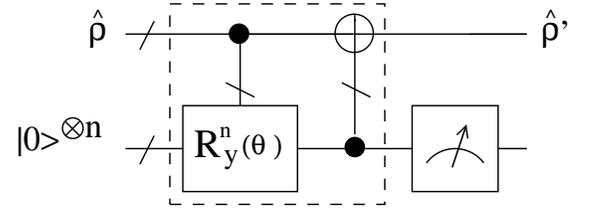}}
  \end{center}
\caption{\footnotesize  Circuit implementation of the generalized amplitude damping channel. All lines (gates) transport (operate on) $n$ qubits.
The marked block has been drawn in detail in Fig. \ref{fig:F311}.\label{fig:F310}} 
\end{figure}

\begin{figure}[h!]
  \begin{center}
    \scalebox{0.6}{%
    \includegraphics*[0,0][428,269]{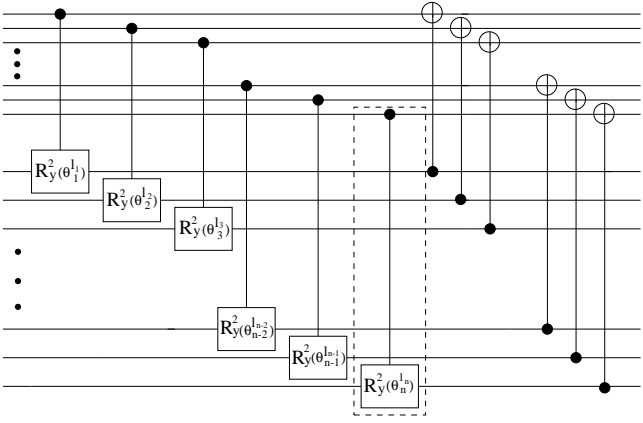}}
  \end{center}
\caption{\footnotesize Enclosed area of the circuit in Fig. \ref{fig:F310}. The upper (lower) lines are $n$ and transport the principal system's (the environment's) qubits.
The $R^{2}_{y}(\theta^{l_{j}}_{j})$ gate (drawn explicitly in Fig. \ref{fig:F312}) is the simple controlled-rotation  gate $R_{y}(\theta^{0}_{j})$ when it is $l_{n}=0$; and 
$R_{y}(\theta^{1}_{j})$ when it is $l_{n}=1$. \label{fig:F311}}
\end{figure}

 For a circuit implementation of the generalized amplitude damping channel we need a circuit that takes the $i^{th}$ computational state and sends it to the
$(i+\mu)^{th}$ one with a probability $p_{i, i+\mu}$. A summarized scheme of such a circuit can be seen in Fig. \ref{fig:F310}. Once again  the symbol ``/'' 
stands for transportation of and operation over $n$ qubits. For example, $R_{y}^{n}(\theta)$ represents a controlled $2n$-qubit gate that applies
a different series of $n$ rotations, to be described below, 
on the $n$ lower qubits (those of the environment, all initialized in the state $\ket{0}$) for  every different state of the $n$ upper ones (those of the principal system)
that work as the control. And the controlled-NOT marked with the symbol
``/'' represents $n$ simple controlled-NOT gates each one acting independently on the $n$ pairs of qubits one from the  principal system and one from the environment. For
the sake of clarity, we display in Fig.  \ref{fig:F311} the enclosed area of the circuit in Fig.  \ref{fig:F310} with every qubit line drawn explicitly. We can see there the
decomposition of the $n$-qubit-pair gate $R_{y}^{n}(\theta)$ into the $n$ two-qubit gates $R^{2}_{y}(\theta^{l_{j}}_{j})$, $0\le j\le n$, which are
not  simple controlled rotations like
those in Fig. \ref{fig:F39} either. Rather, they are  gates that apply the simple rotation operator around the $y$-axis
 in an angle $\theta^{l_{j}}_{j}$,
$\hat{R}_{y}(\theta^{l_{j}}_{j})$, to the environment's $j^{th}$  qubit when the principal system's $j^{th}$  qubit is in  state $\ket{l_{j}}$, where $l_{j}$ can take the
values $0$ or $1$. Thus there are altogether $2n$ angles ($2$ for every one of the $n$ two-qubit gates $R^{2}_{y}(\theta^{l_{j}}_{j})$) to
choose in order to univocally specify this 
circuit's action.
\par Writing the $i^{th}$, $0\le i\le N-1=2^{n}-1$, computational state in base $2$, $\ket{i_{2}}=\ket{l_{1}\times\ket{l_{2}}... \ \
...\ket{l_{n-1}}\otimes\ket{l_{n}}}\equiv\ket{l_{1}l_{2}... \ \ ...l_{n-1}l_{n}}$ (where the subindex $2$ under the $i$ stands for
``binary representation of'') , it is straightforward to see \cite{tesis} that after application of the entire circuit the state is
taken to $\ket{l_{1}\oplus s_{1} \ \ l_{2}\oplus s_{2} ... \ \ ...l_{n-1}\oplus s_{n-1}\ \ l_{n}\oplus s_{n}}$ (where ``$\oplus$'' is the
bit-to-bit sum that must be done adding modulo $2$ every bit from a number with its corresponding bit from the other and  where $s_{j}$
can only take the values $0$ or $1$ too) with a probability 
\[
  p_{i_{2}, i_{2}\oplus s_{1}s_{2}... \ \ ...s_{n-1}s_{n}}=f(s_{1},\theta^{l_{1}}_{1})f(s_{2},\theta^{l_{2}}_{2})... 
\]
\begin{equation}
  \label{eq:probabilidades}
  ...f(s_{n-1},\theta^{l_{n-1}}_{n-1})f(s_{n},\theta^{l_{n}}_{n})\ \ ,
\end{equation}
being
\begin{equation}
  f(s_{j},\theta^{l_{j}}_{j})=
  \left\{
  \begin{array}{ll}
     \cos^{2}\Big({\frac{\theta^{l_{j}}_{j}}{2}}\Big)  , \ \ \text{for} \ \ s_{j}=0 \ \ , \\
     \sin^{2}\Big({\frac{\theta^{l_{j}}_{j}}{2}}\Big)  , \ \ \text{for} \ \ s_{j}=1 \ \  .
  \end{array}
  \right.
\end{equation}  
So for every $j$ there are $4=2^{2}$ possible values which give us altogether $(2^{2})^{n}=(2^{n})^{2}=N^{2}$ different combinations of cosines and sines in
(\ref{eq:probabilidades}), that is exactly what we need to specify the $N^{2}$ transition probabilities $p_{i i+\mu}$. Thus making the following identification for the binary
representation of the number $\mu$, $\mu_{2}=s_{1}s_{2}... \ \ ...s_{n-1}s_{n}$, we get the desired action
: a circuit that takes every one of the $N$ computational states
$\ket{i}$ of the principal system to a superposition of the  $N$ states $\ket{i+\mu}$ with probabilities $p_{i_{2}, i_{2}+\mu_{2}}$ given by Eq.
(\ref{eq:probabilidades}) (which are in turn
controlled choosing the $2n$ available rotation angles) that, after application of the measurement, becomes an incoherent superposition of them.
\begin{figure}[h!]
  \begin{center}
    \scalebox{0.5}{%
    \includegraphics*[0,0][251,195]{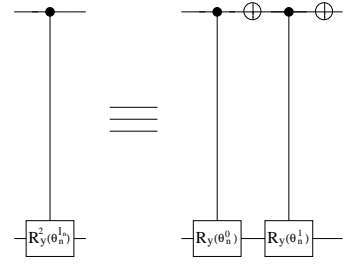}}
  \end{center}
\caption{\footnotesize Detailed decomposition of the gate $R^{2}_{y}(\theta^{l_{j}}_{j})$ enclosed in Fig. \ref{fig:F311} in terms of simple controlled-rotation and NOT gates. If the
control is in state $\ket{0}$, the applied rotation is in an angle $\theta^{0}_{j}$ ($\hat{R}_{y}(\theta^{0}_{j})$); and  when it is in state $\ket{1}$, the  rotation is
in an
angle $\theta^{1}_{j}$ ($\hat{R}_{y}(\theta^{1}_{j})$). \label{fig:F312}} 
\end{figure}
\subsection{The spectra.}
   \label{spectra}
\begin{figure}[hr!]
  \begin{center}
    \psfrag{0.5}{$0.5$}
    \psfrag{-0.5}{$-0.5$}
    \scalebox{0.73}{%
    \includegraphics*[95,421][440,764]{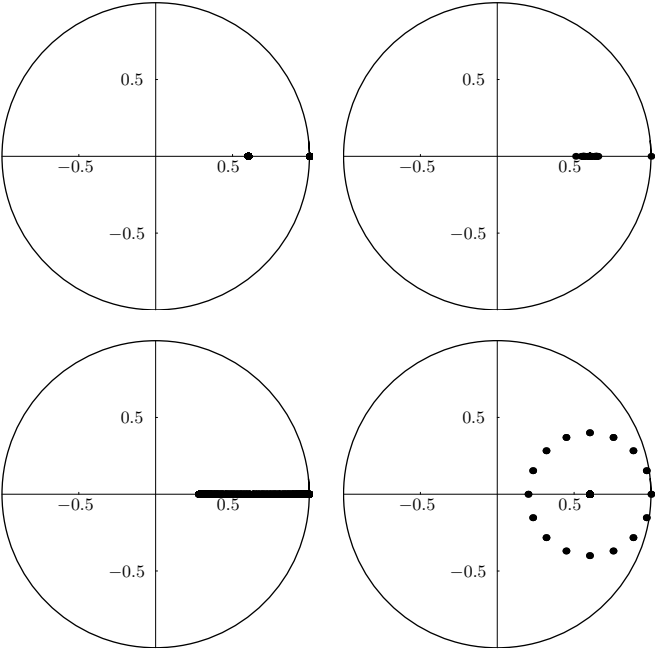}}
  \end{center}
\caption{\footnotesize Numerical spectra of the maps (clockwise) $\$_{DC}$ and $\$_{PDC}$, $\$_{PDC}$ for a random choice of the
coefficients of $C_{ij}$, $\$_{L_{102}}$, and $\$_{ADC}$ in the complex plane. The parameters are $N=32$ and $\epsilon=0.4$. \label{fig:F43}} 
\end{figure}

We now turn to an analysis of the spectra of these superoperators, which in all cases except for the ADC can be obtained analytically. (See Fig. \ref{fig:F43}.) The top
left-hand graph corresponds to the spectra of the generalized depolarizing channel and the
  generalized phase damping channel: in both cases the eigenvalues are at $\lambda= 1$ and $\lambda=1-\epsilon $,  but their degeneracies differ.
  For the case of $ \$_{DC} $ the unit eigenvalue  (corresponding to the eigenoperator $\hat{I}$) is non-degenerate and the
  other  eigenvalue has degeneracy $N^{2}-1$. For  the case of $\$_{PDC}$
   the degeneracy of the eigenvalue $1$ is $N$ (corresponding to the $N$  diagonal projectors $\hat{P}_{ii}$), 
   while that of the eigenvalue $1-\epsilon$ is $N^{2}-N$ (corresponding to the non diagonal projectors $\hat{P}_{ij}$, $i\neq j$, which contain the coherences of the density matrix.
  This $(N^{2}-N)$-fold degeneracy at  $1-\epsilon$ depends on the choice
$C_{ij}=\delta_{ij}$ for the coefficients in (\ref{eq:PDC}). For random coefficients the
   degeneracy is slightly broken, as shown in the top right-hand plot. In the bottom right-hand graph the spectrum of  the generalized phase damping channel on a line for the choice $n_{1}=1$,
      $n_{2}=0$ and $n_{3}=2$ is shown; there the degeneracy has been broken even more strongly and $N$ out of the $N^{2}-N$ that were initially at $1-\epsilon$ have been
   spread along the circumference of center $1-\epsilon$ and radius $\epsilon$ in pairs (the degeneracy of these $N$ eigenvalues depend
 on the parameters $n_{1}$, $n_{2}$ and   $n_{3}$). The bottom left-hand graph shows the spectrum of the generalized amplitude damping channel;
  here the previous degeneracies have been completely broken (the eigenvalue $1$, of eigenoperator $\hat{I}$, is simple and all the
  others are at most doubly degenerate.
\par In view of the spectra we can now understand some features of the action of these maps in phase space. The fact that the microcanonical distribution 
is the  only invariant state of $\$_{DC}$ is due to $\hat{I}$ being its only eigenoperator with eigenvalue $1$. In contrast, 
the invariant subspace of the generalized phase damping channels is $N$-dimensional, spanned by the $N$ projectors $\hat{P}_{ii}$ on  the
pointer states. On the other hand, $\$_{ADC}$, 
when expressed in the chord basis in terms of translations, does not take a diagonal form as was the case for $\$_{DC}$ and $\$_{DC}$. This is now not surprising due to its
spectrum's very weak degeneracy \cite{comentario4}.
\section{Noisy unitary evolution}
   \label{Noise}
\par
As was pointed out in the introduction apart from studying the decoherence models themselves the other  goal of this work is to
characterize their effects on unitary maps, to study the noisy unitary evolution. For this aim we will model the noisy  
evolution with two-stage superoperators as is done in the literature \cite{Yo, garma, garma2, nonn, tesis}. That is, given a unitary 
superoperator $U$, such that $\hat{\rho}'\equiv U(\hat{\rho})=\hat{U}\hat{\rho}\hat{U}^{\dagger}$, we will consider a map composed of a
unitary step plus a pure decoherent step described by a superoperator $\$$ (which, for our case, will be one of the studied channels), 
$\$_{\circ} U$, such that $\hat{\rho}'\equiv (\$_{\circ}U)(\hat{\rho})\equiv \$(U(\hat{\rho}))$. In this approximation the total evolution occurs by alternating unitary and noisy steps. 
\subsection{Noisy Grover's algorithm.}
   \label{NoisyGrover}
\par
The first unitary map we will study is Grover's search algorithm  which has no classical analog but is one of the simplest examples 
of algorithms whose quantum versions are more efficient than the classical ones \cite{preskill, chuang}. The algorithm consists of the successive
application of the unitary operator $\hat{U}_{G}$ on the state $\ket{\psi}$ initialized as the uniform superposition of all $N$ computational states (once again taken
\begin{figure*}[ht!]
  \begin{center}
    \scalebox{1.2}{%
    \includegraphics*[102,661][530,771]{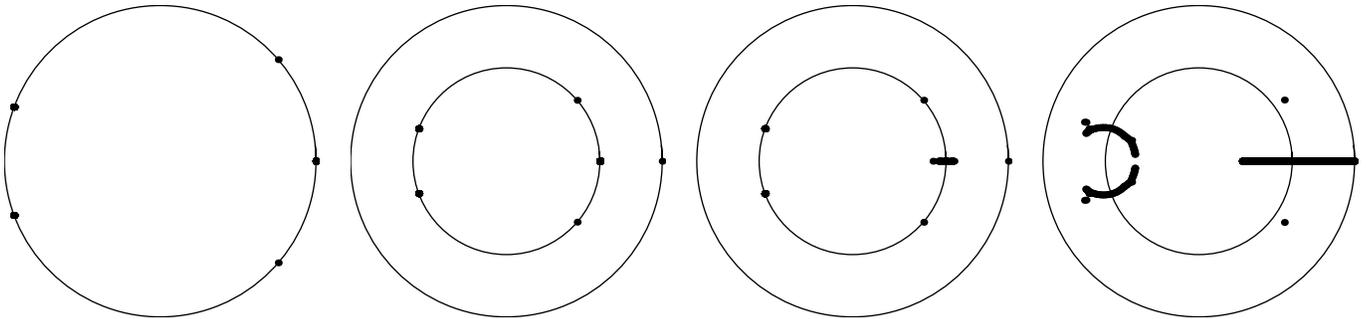}}
  \end{center}
\caption{\footnotesize Numerical spectra of the maps (from left to right) $U_{G}$, $(\$_{DC})_{\circ}(U_{G})$, $(\$_{PDC})_{\circ}(U_{G})$ and 
$(\$_{ADC})_{\circ}(U_{G})$ in the complex plane for $N=32$ and $\epsilon=0.4$. All the eigenvalues of the unitary map $U_{G}$ lie on the unitary circumference. 
$\$_{DC}$ contracts the spectrum radially without breaking any degeneracy except for the eigenvalue $1$, $\$_{PDC}$ also contracts it radially leaving the eigenvalues'
phases intact but inducing a slight splitting of some eigenvalues. For last, $\$_{ADC}$ breaks all degeneracies strongly in the radial and angular directions. See text.\label{fig:F5354}} 
\end{figure*} 
as the position states),
$\ket{\psi}=\frac{1}{\sqrt{N}}\sum_{q=0}^{N-1}\ket{q}$ (which is equivalent to the first element of the momentum basis, $\ket{\psi}=\ket{p=0}$). It is designed so that 
after an optimal number of iterations $T\approx\frac{\pi}{4}\sqrt{\frac{N}{M}}$ the
$M$ marked elements $\ket{w}$ will have a large probability in the computational basis.
 In turn, the operator $\hat{U}_{G}$ can be decomposed 
as $\hat{U}_{G}=\hat{U}_{\psi}\hat{U}_{O}$. $\hat{U}_{O}$ represents the call to an oracle whose only capacity is to distinguish the marked items from the rest.
$\hat{U}_{\psi}$ is the inversion-about-the-mean operator that contains no information about the marked items and can be written as
$\hat{U}_{\psi}=\hat{I}-2\ket{\psi}\bra{\psi}|$.
\par
\begin{figure}[h!]
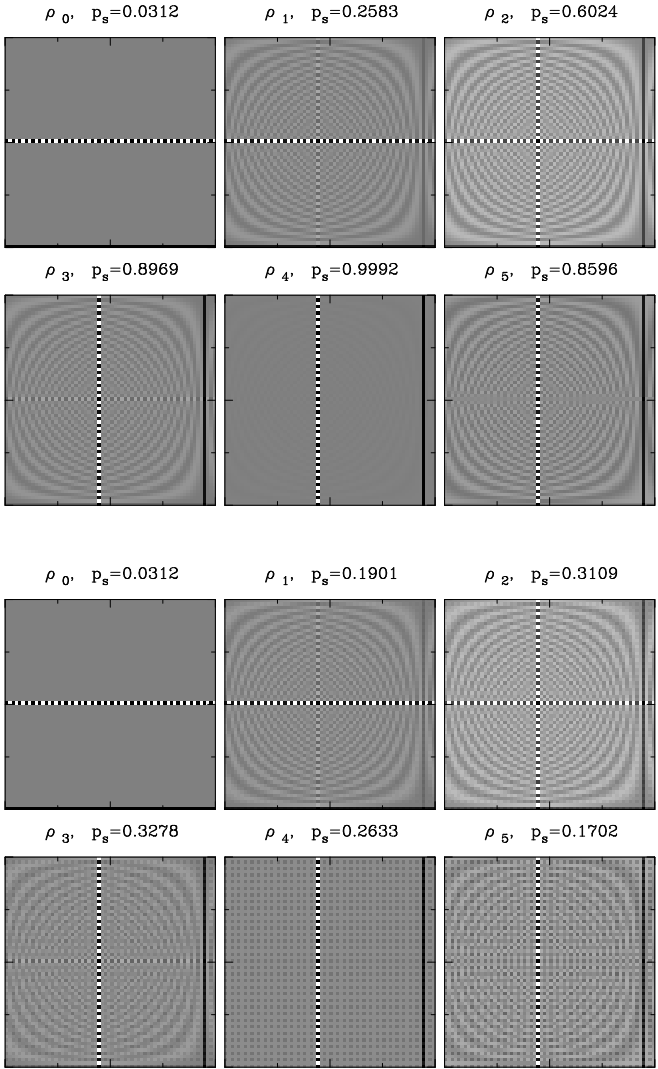

\begin{center} 
    \scalebox{0.5}{%
    \includegraphics*[59,88][554,469]{F52s.ps}}
\end{center}
\begin{center}
    \scalebox{0.5}{%
    \includegraphics*[59,88][553,469]{F55s.ps}}
\end{center}
\caption{\footnotesize (Above) Unitary Grover's algorithm in phase space. After only $4$ iterations the initial momentum state $\ket{p=0}$ is practically transformed into
a position state localized precisely at the position of the marked item. From then on the success probability $p_{s}$ starts decreasing again (see text). (Bottom) Grover's
algorithm perturbed with the depolarizing channel, the parameters are the same as those used in Fig. \ref{fig:F55,F56&F57}. \label{fig:F52}}
\end{figure}

 In Fig. \ref{fig:F52} (top) we can see the action of the algorithm in phase space for $N=32$, $M=1$  and $w=30$ (only one marked item, the penultimate one). Initially
the state of the quantum computer is the momentum state $\ket{p=0}$ uniformly distributed over all coordinate states. After every iteration  the probability gradually concentrates on a position eigenstate centered on the marked  $\ket{w}$. We can also see
the growth of the success probability $p_{s}$ (initially equal to $\frac{1}{32}=0.0312$) in such a way that if a measurement is performed right after the fourth iteration 
($T\approx\frac{\pi}{4}\sqrt{\frac{N}{M}}=\frac{\pi}{4}\sqrt{\frac{32}{1}}\approx 4$) we will have more than a $99  $ per cent
chance of getting $\ket{w}$ as the result. Note 
that in the fifth iteration this probability starts decreasing again (that is why it is so important to perform the measurement right at the optimal iteration).
\par While the action of the unitary algorith in phase space was already studied in \cite{Miquel, Bianucci}; the spectrum of the map provides a  perspective 
from a different point of view. It is  possible to write the initial state of the quantum computer as a combination of a uniform superposition
of the $N-M$ non marked computational states, $\ket{\alpha}=\frac{1}{\sqrt{N-M}}\sum_{q\in N-M}\ket{q}$, and a uniform superposition of the $M$ marked ones, 
$\ket{\beta}=\frac{1}{\sqrt{M}}\sum_{w\in M}\ket{w}$; that is,
$\ket{\psi}=\sqrt{\frac{N-M}{N}}\ket{\alpha}+\sqrt{\frac{M}{N}}\ket{\beta}$. The particular thing about these two vectors is that they span a $2$-dimensional invariant
subspace of $\hat{U}_{G}$ in which its matrix representation is just a $2\times 2$ rotation matrix in an angle $\theta$, such that $\sin{\theta}=\frac{2\sqrt{(N-M)M}}{N}$.
This leads us to the well known geometrical representation of the algorithm \cite{chuang}: the initial vector $\ket{\psi}$ is
rotated in the $\alpha\beta$-plane in an angle $\theta$ per iteration up to the optimal iteration $T$ when the projection of $\ket{\psi}$ on  $\ket{\beta}$ (and thus the
success probability $p_{s}$) is maximized. If the iteration process is kept on,  the vector   $\ket{\psi}$ continues to rotate and $p_{s}$ starts to decrease again and so
forth. The eigenvalues of this rotation matrix are $\lambda_{\pm}=e^{\pm i\theta}$ with eigenvectors $\ket{\lambda_{\pm}}=\frac{1}{\sqrt{2}}(\ket{\alpha}\pm i\ket{\beta})$.
These two eigenvalues of the operator $\hat{U}_{G}$ give rise in turn to four eigenvalues of the map $U_{G}$ ($\equiv\hat{U}_{G}\odot\hat{U}_{G}^{\dagger}$):
 $e^{\pm 2i\theta}$ and $e^{\pm i0}=1$, of eigenoperators $\ket{\lambda_{\pm}}\bra{\lambda_{\pm}}|$ and $\ket{\lambda_{\pm}}\bra{\lambda_{\mp}}|$ respectively, which
 analogously span a $4$-dimensional invariant subspace for  $U_{G}$. As far as this subspace is concerned,
 the  action of the map would be exactly the same regardless of the eigenvalues corresponding to operators outside the subspace. If the density operator is
 initialized in the state $\ket{\psi}\bra{\psi}|$ it 
never leaves the subspace. Therefore, the algorithm would work equally well no matter what the other eigenvalues are. We shall now see that when we consider the map composed by $U_{G}$
and some of our channels the subspace spanned by $\ket{\lambda_{\pm}}\bra{\lambda_{\pm}}|$ and $\ket{\lambda_{\pm}}\bra{\lambda_{\mp}}|$ is no longer invariant. Thus the
relationship of these four principal eigenvalues and all the others begins to be relevant for the search's  efficiency. 
\par In Fig. \ref{fig:F5354} we can see the numerical spectra of the maps (from left to right) $U_{G}$,$(\$_{DC})_{\circ}(U_{G})$, $(\$_{PDC})_{\circ}(U_{G})$ and 
$(\$_{ADC})_{\circ}(U_{G})$ for $N=32$ and $\epsilon=0.4$. The first circle shows the eigenvalues of the unitary map by itself. The eigenvalue $1$ and those
in the second and third quadrants are very degenerate while those in the first and fourth ones are simple. The latter are the two eigenvalues corresponding to the
eigenoperators  $\ket{\lambda_{\pm}}\bra{\lambda_{\pm}}|$, while those corresponding to $\ket{\lambda_{\pm}}\bra{\lambda_{\mp}}|$ are concentrated at $1$. 
\begin{figure}
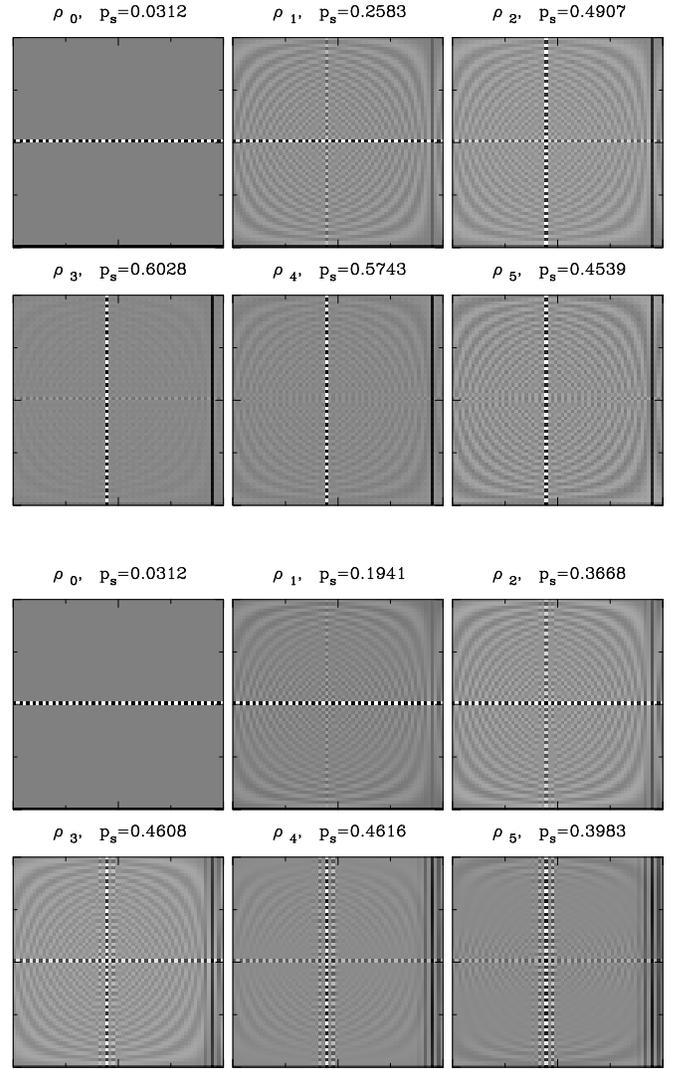

\begin{center}
   \scalebox{0.5}{%
    \includegraphics*[59,88][553,469]{F56s.ps}}
\end{center}
\begin{center}
    \scalebox{0.5}{%
    \includegraphics*[59,88][553,469]{F57s.ps}}
\end{center}
\caption{\footnotesize DPS representation of the noisy Grover's algorithm ($N=32$ and $\epsilon=0.3$). The algorithm's efficiency is altered both in a
global decrease of the success probability $p_{s}$ with respect to the unitary case and in the value of the optimal iteration $T$. The algorithm is more resistant before
the phase damping channel (top) and the amplitude damping channel (bottom) than before the depolarizing channel (Fig. \ref{fig:F52} bottom). See text.\label{fig:F55,F56&F57}}
\end{figure} 
\par In \cite{tesis} it
was shown that the action of $\$_{DC}$ on the spectrum of unitary maps is to contract the spectrum uniformly in the radial direction by a factor $1-\epsilon$ except for the
eigenvalue of the eigenoperator $\hat{I}$ that is left unchanged at $1$.  If an eigenoperator of the unitary map is orthogonal to $\hat{I}$ then it is also left intact with its associated eigenvalue
reduced by the factor $1-\epsilon$; but if it is not, then it is its orthogonal complement to $\hat{I}$ what is left as an eigenoperator of the composed map with the
associated eigenvalue reduced by $1-\epsilon$ too. And indeed, this is what can be seen in the second graph where all eigenvalues have been projected radially from the
unit circumference to the circumference of radius $1-\epsilon$ (also drawn explicitly) except for one simple eigenvalue that has been left at $1$, that of the identity operator. As
forseen before, the space spanned by $\ket{\lambda_{\pm}}\bra{\lambda_{\pm}}|$ and $\ket{\lambda_{\pm}}\bra{\lambda_{\mp}}|$ is no longer invariant under the action the composed map,
for while $\ket{\lambda_{\pm}}\bra{\lambda_{\mp}}|$ are traceless operators (and thus orthogonal to $\hat{I}$), $\ket{\lambda_{\pm}}\bra{\lambda_{\pm}}|$ are not. So 
$\ket{\lambda_{\pm}}\bra{\lambda_{\mp}}|$ are still eigenoperators of the composed map now with eigenvalue $1-\epsilon$, but not $\ket{\lambda_{\pm}}\bra{\lambda_{\pm}}|$ 
whose orthogonal complements to $\hat{I}$ are multiplied by $(1-\epsilon)e^{\pm 2i\theta}$ and their projections on $\hat{I}$ by $1$. We conclude this way that in this case there
is a competition iteration by iteration between the tendency of the algorithm to take the initial state to the state of interest $\ket{w}\bra{w}|$ and the tendency of the noise 
to take it to $\frac{\hat{I}}{N}$.
\par In the third graph we can see how some degeneracies of those eigenvalues that were at $1$ and now are around the point $1-\epsilon$ have been broken. 
But the most important thing is that the eigenvalues left at $1$ are now $N$, those corresponding not to $\hat{I}$ but to any diagonal projector $\hat{P}_{ii}$ (or
alternatively, to
any diagonal matrix). The loss of information is much less abrupt in this process than in the previous one, so we expect the algorithm to work better when perturbed by the
phase damping channel than by the depolarizing channel.
\par In the fourth graph we can see how  $\$_{ADC}$ breaks all degeneracies strongly in the radial and angular directions. This time there is only one eigenvalue left at
$1$, that of $\hat{I}$ as in the case of $\$_{DC}$. But, in contrast, there are many eigenvalues  forming a quasi-continuum close to $1$ and, what is most important, 
 two of the principal eigenvalues (in the first and fourth quadrants) are not reduced as much as in the previous cases (they  are outside the $(1-\epsilon)$-radius circle). So it is
reasonable to expect a better functioning of the algorithm for this case too. 
\par All the conclusions drawn above from the spectra can be confirmed in phase space comparing the unitary (Fig. \ref{fig:F52} top) and the noisy cases (Fig. \ref{fig:F52}
bottom and Fig. \ref{fig:F55,F56&F57}). As in the upper plot of Fig. \ref{fig:F52}, the lower is the Wigner function
representation of the evolution of the computer's state initialized as $\hat{\rho}=\ket{\psi}\bra{\psi}|=$ $\ket{p=0}\bra{p=0}|$  but under the action 
of the composed map $(\$_{DC})_{\circ}(U_{G})$. We can see how all success probabilities are drastically reduced with respect to the unitary algorithm. But another important aspect to notice is that the highest success probability is now attained at the third iteration. Evidently, at the fourth iteration (the former optimal iteration) the tendency from the noise
to take the computer to the completely mixed state (evident from the gradual appearance of the chessboard-like pattern) becomes dominant over the tendency from the
algorithm to find the chosen item. Fig. \ref{fig:F55,F56&F57} top corresponds to $(\$_{PDC})_{\circ}(U_{G})$. This time the maximal success probability is
attained also at the third iteration instead of the fourth one, but the values obtained for the probabilities are about two times as high as those for the depolarizing channel case in accordance to
what was anticipated in the discussion about the spectra. The lower graph of Fig. \ref{fig:F55,F56&F57} corresponds to $(\$_{ADC})_{\circ}(U_{G})$. Here the probabilities are a little bit lower than
for $(\$_{PDC})_{\circ}(U_{G})$ but certainly still much higher than for $(\$_{DC})_{\circ}(U_{G})$ and the optimal iteration is the fourth one as in the unitary case, again
in accordance with what predicted from the spectra. On the other hand the lines parallel to that one of the marked item observed in the last iterations are nothing but the
effect of the horizontal spreading $\$_{ADC}$ provokes.
\subsection{Noisy cats and bakers.}
   \label{NoisyMaps}
\par The second family of quantum maps we shall study are the quantum cat maps studied by Hannay and Berry  \cite{hannay}. They quantize the classical motion represented by  the  transformations from the torus onto itself given by the symplectic matrix 
\begin{equation}
  M=
  \left(\begin{array}{cc}
		2\beta & -1 \\
		1-4\alpha\beta & 2\alpha
	   \end{array}
   \right) ,\ \ 
   \left(\begin{array}{cc}
		q^{,} \\
		p^{,}
	   \end{array}
   \right) =
   M \left(\begin{array}{cc}
		q \\
		p
	   \end{array}
   \right);
   \label{Matrizmapas}
\end{equation} 
where $\alpha$ and $\beta$ are integers and $(q^{,},p^{,})$ are the coordinates of the transformed point $(q,p)$. 
The dynamics implied by this map is determined by the eigenvalues of $M$. A hyperbolic map is obtained by choosing $\alpha=\beta=1$. In this case we obtain $\lambda_{H_{\pm}}=2\pm\sqrt{3}$, with eigenvectors $(q_{\pm},p_{\pm})=(1,\pm\sqrt{3})$ along the stable and unstable directions. For 
 $-\alpha=\beta=1$ we obtain $\lambda_{E_{\pm}}=\pm i$ with no real eigenvectors, and the map for this case is  an elliptic rotation in phase space, whose dynamics is completely regular. The parabolic case, representing a phase space shear, can be obtained (for example)  with $\alpha=0$ and $\beta=1$;

\par The quantum propagator for this classical maps is given by the unitary operator $\hat{U}_{C}$, whose matrix representation in the discrete position basis is
\[
  U_{C}(q^{,},q)\equiv\bra{q^{,}}|\hat{U}_{C}\ket{q}
\]
\begin{equation}
 =\frac{1}{\sqrt{N}}e^{-\frac{2\pi}{N}iF_{1}(q^{,},q)}=\frac{1}{\sqrt{N}}e^{-\frac{2\pi}{N}i(\alpha
  q^{,^{2}}-q^{,}q+\beta q^{2})}.
   \label{Propagmapas}
\end{equation} 
 where $N$ as usual stands for the dimension of the Hilbert space and
$F_{1}(q^{,},q)=\alpha q^{,^{2}}- q^{,}q +  \beta q^{2}$ is the generating function  that plays  a fundamental role in the quantization procedure
\cite{hannay}.
\par Let us now introduce the last map we shall study, the baker's map. This is one of the simplest maps displaying strongly chaotic behaviour and, in spite of its
simplicity, it possesses a very rich dynamics both in its classical and quantum versions. The map is is an area-preserving transformation defined in the $[0,1]\times[0,1]$
phase space square (the torus with periodic boundary conditions) as
\[
  q^{,}=2q-[2q],
\]
\begin{equation}
  p^{,}=\frac{1}{2}(p+[2q]);
  \label{pana}
\end{equation}
where the square brackets symbolize the integer part of the number between them. The transformation has a very simple geometrical interpretation, as a ``stretching''
step followed by a ''cutting'' step, as a baker rolling a dough. The map is uniformly hyperbolic with a single Lyapunov exponent $\gamma_{P}=ln(2)$. Moreover, at every
point the stable and unstable manifolds are parallel to the coordinate axes ($(q_{B_{-}},p_{B_{-}})=(1,0)$ and $(q_{B_{+}},p_{B_{+}})=(0,1)$. To quantize it we follow
the quantization procedure of Balasz and Voros \cite{BalazsVoros}, which yields the following unitary quantum propagator:
\begin{equation}
  \hat{U}_{P}=\hat{U}_{F_{N}}^{\dagger}
  \left(\begin{array}{cc}
		\hat{U}_{F_{N/2}} & 0 \\
		0 & \hat{U}_{F_{N/2}}
	   \end{array}
   \right) ; 
\end{equation} 
where $\hat{U}_{F_{N}}$ is the change of basis matrix from the position to the momentum basis (that is, the discrete Fourier transform) and whose matrix elements in
the position representation are $U_{F_{N}}(q^{,},q)=\bra{q^{,}}|\hat{U}_{F_{N}}\ket{q}=\frac{1}{\sqrt{N}}e^{-\frac{2\pi}{N}iq^{,}q}$.
\par We quantify the action of the noise on the map through the linear entropy $S$, defined as $S\equiv
 -ln(Tr(\hat{\rho}^{2}))$.  The minimum value that this quantity 
takes is $S_{Min}=0$, which corresponds to a pure state; while the maximum is $S_{Max}=ln(N)$ and corresponds to the completely mixed state $\frac{\hat{I}}{N}$.
\par
\begin{figure}[ht!]
\begin{center} 
    \scalebox{.8}{%
    \includegraphics*[99,507][405,704]{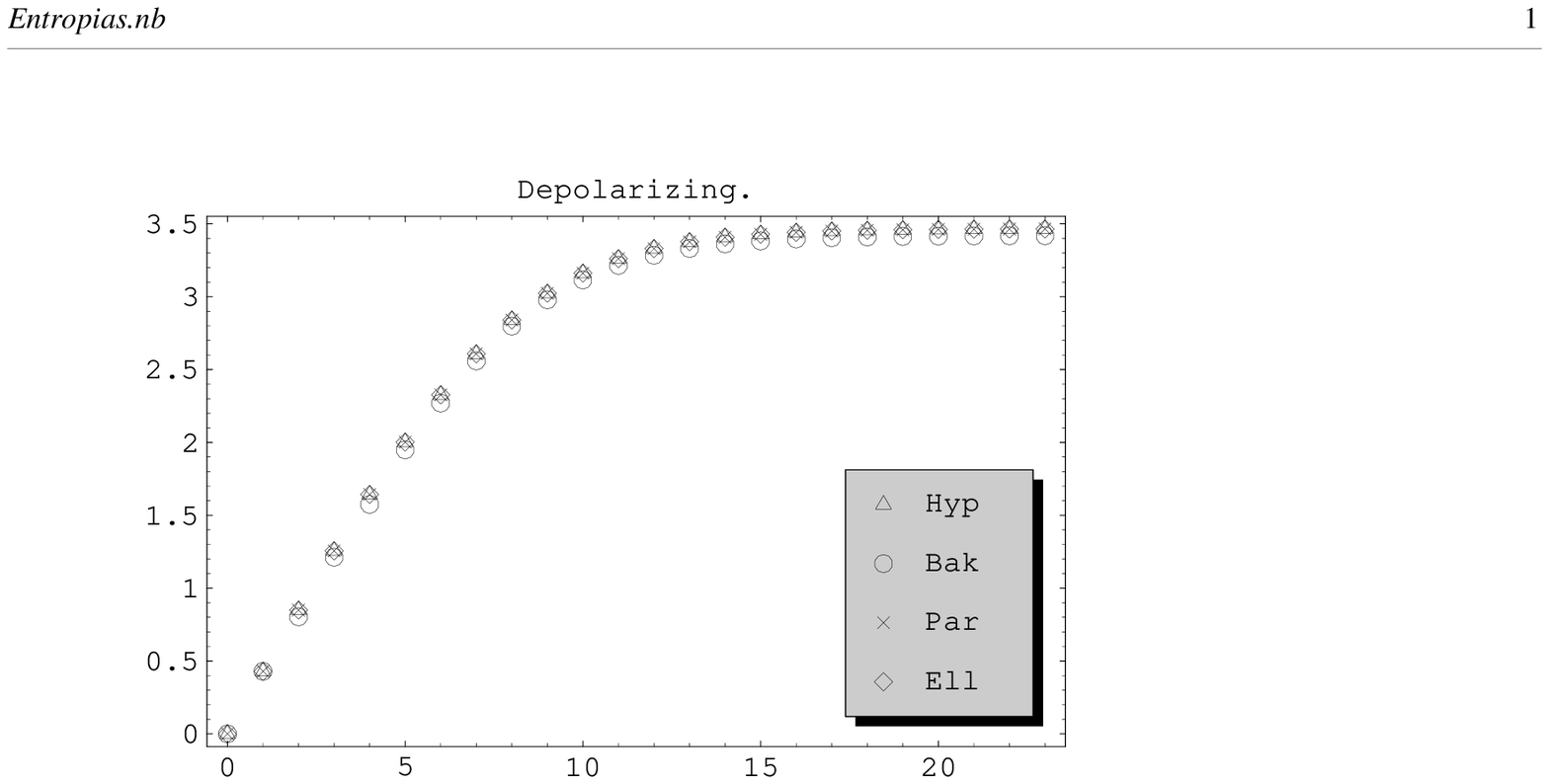}}
\end{center}
\begin{center} 
    \scalebox{.8}{%
    \includegraphics*[99,514][405,710]{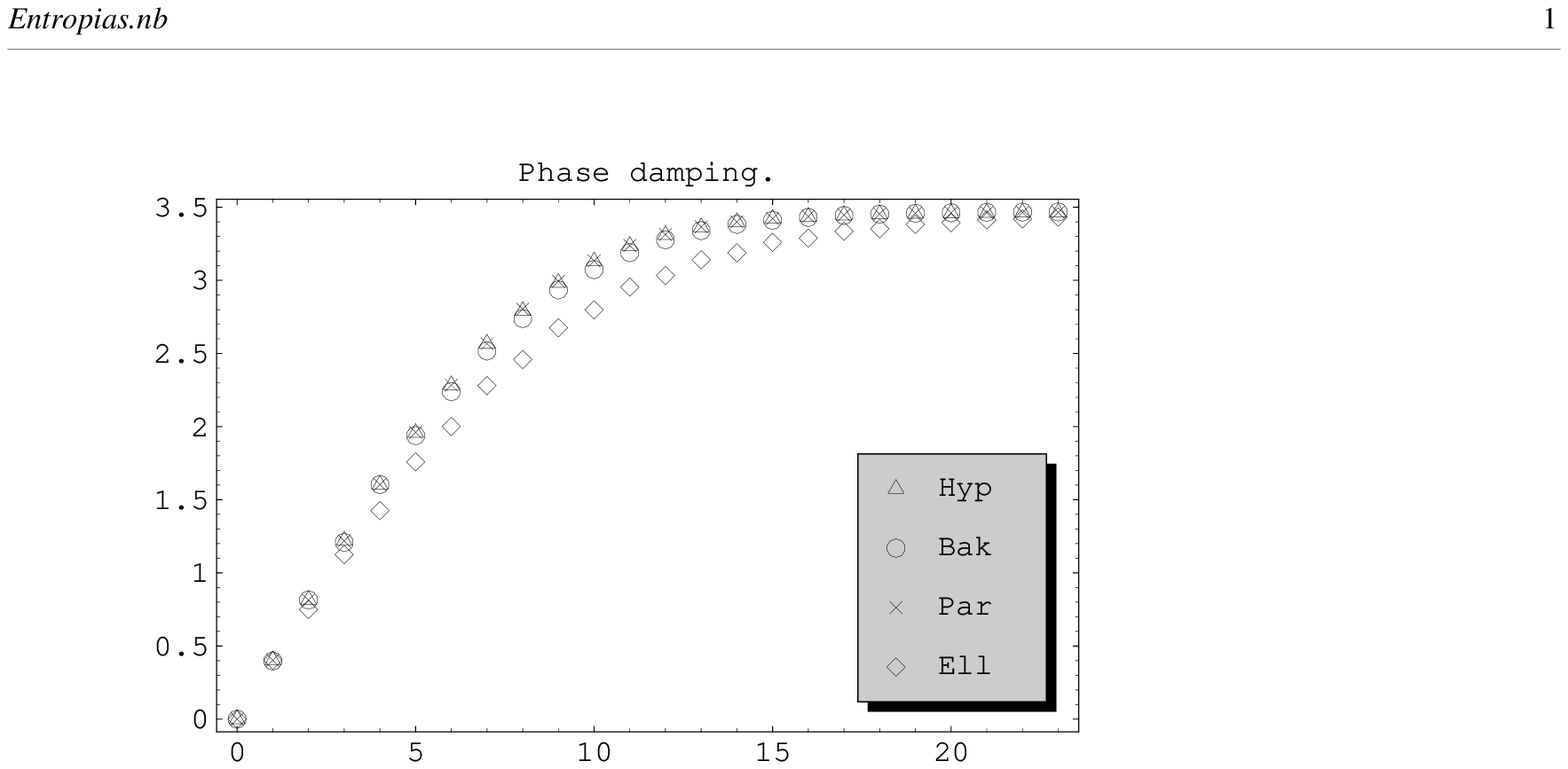}}
\end{center}
\begin{center} 
   \scalebox{.8}{%
    \includegraphics*[99,510][405,711]{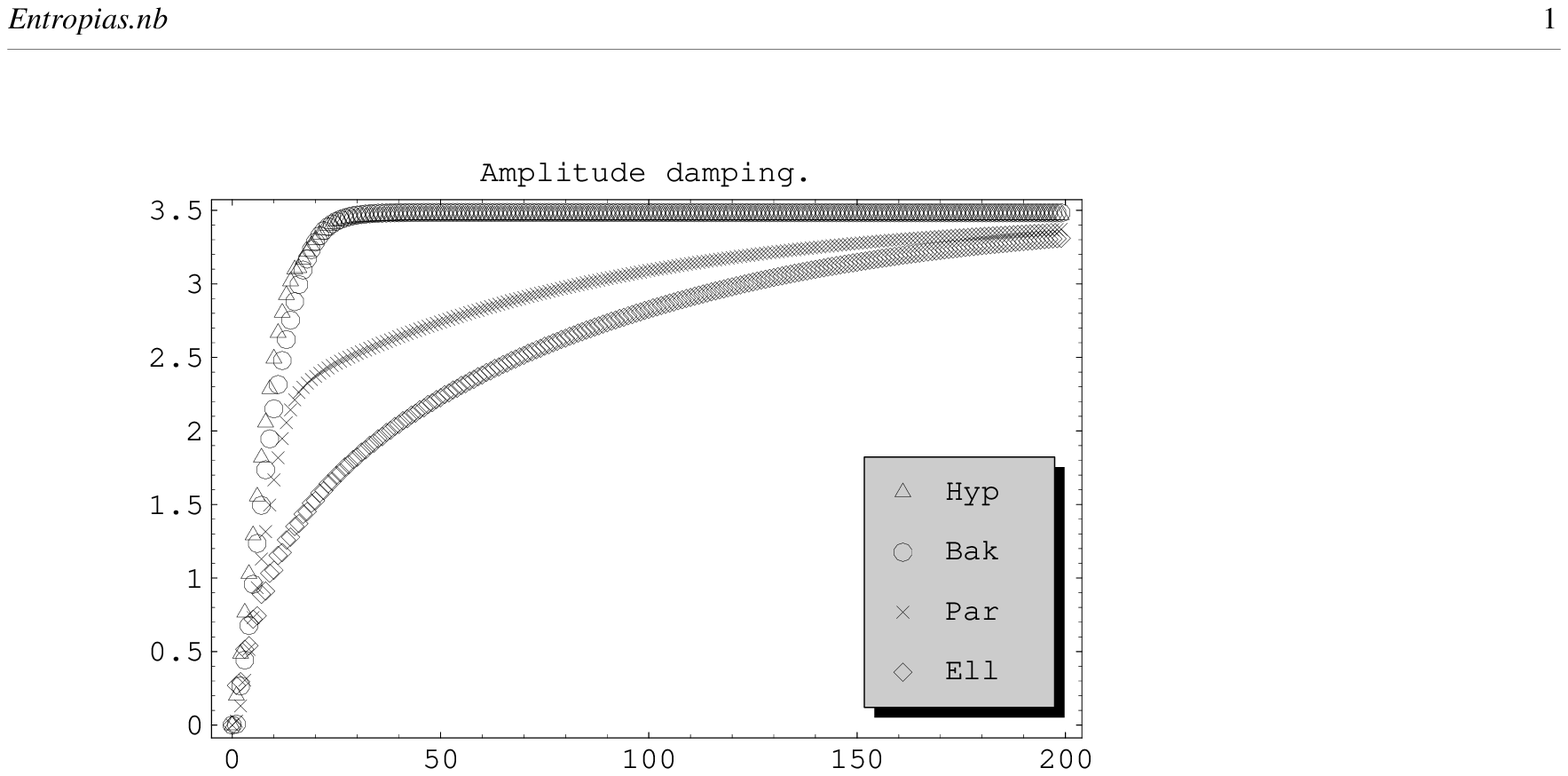}}
\end{center}
\caption{\footnotesize Linear entropy $S$ as a function of the number of iterations for $\$_{DC}$ (top), $\$_{PDC}$ (center) and $\$_{ADC}$ (bottom) composed with the
four unitary maps; $N=32$ and $\epsilon=0.2$. The computer has been initialized in a coherent state. For the case of  $\$_{ADC}$ the number of iteration has been taken
much greater so as  to fully appreciate the evolution of $S$ until it reaches saturation (note the difference in scales). The behaviour of the entropy depends on both the
degeneracy of the decoherent superoperator and the degree of complexity of the associated classical map (see text).\label{tresruid4maps}}
\end{figure}

 In Fig. \ref{tresruid4maps} we have plotted the evolution of the linear entropy of an initially coherent state as a function of the number of iterations of the three
generalized channels composed with the four quantum maps. In all cases we have taken $N=32$ and $\epsilon=0.2$.
\par The uppermost plot corresponds to $\$_{DC}$, the completely depolarizing and most degenerate superoperator. We can observe there how it induces the same entropy
growth for all four maps: there exists first a linear regime and then an asymptotic tendency to saturation at the value $S_{Max}=ln(N)=ln(32)\approx 3.47$. So, as far as $S$
is concerned, it is as though the application of $\$_{DC}$ ``erased'' the dynamics imprinted on the state by the previous unitary step. And, in view of the discussion of
the last paragraph of Sec. \ref{GDC}, that is not much of a surprise; because we know this map's action can be understood as leaving the state intact, with probability
$1-\epsilon$, or taking it to $\frac{\hat{I}}{N}$, with probability $\epsilon$, no matter what the initial state was (or, in particular, no matter what the applied unitary
map was).
\begin{figure*}[ht!]
\begin{center} 
    \scalebox{1.32}{%
    \includegraphics*[101,360][494,761]{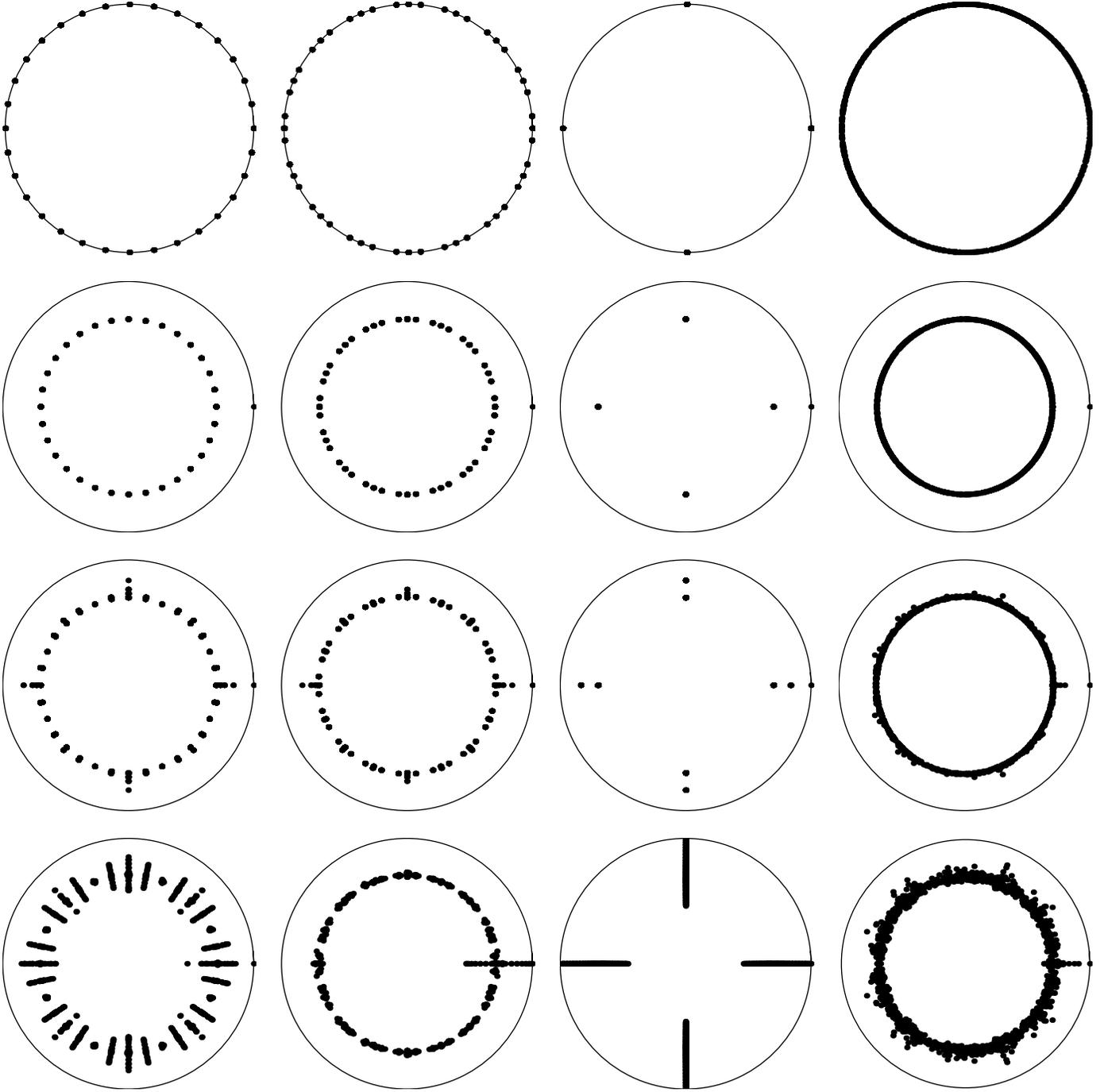}}
\end{center}
\caption{\footnotesize Upper row: (from left to right) numerical spectra of the unitary maps $U_{HC}$, $U_{PC}$, $U_{EC}$ and $U_{B}$ The spectra seen in the three lower
rows correspond to (in descending order) $\$_{DC}$, $\$_{PDC}$  and $\$_{ADC}$ composed with the same maps ($N=32$ and $\epsilon=0.2$). \label{todoslosespectros}}
\end{figure*} 
\par In the central plot we can see the evolution of the entropy but this time for the composition of $\$_{PDC}$ (whose spectrum is less degenerate than $\$_{DC}$'s) with
the four maps. In this case the situation is slightly different, for the curves corresponding to $(\$_{PDC})_{\circ}(U_{B})$, $(\$_{PDC})_{\circ}(U_{H})$ and 
$(\$_{PDC})_{\circ}(U_{P})$ display pretty much the same behaviour as before: but on the contrary, for the case of $(\$_{PDC})_{\circ}(U_{E})$ a little slower entropy
growth is observed.
\par Finally, in the lowermost graph we can see a completely different situation for $\$_{ADC}$, the superoperator with the least degenerate spectrum we have studied. The
curves corresponding to $(\$_{ADC})_{\circ}(U_{B})$ and $(\$_{ADC})_{\circ}(U_{H})$ are almost identical and, before saturation, they present a linear growth regime as
before but with a considerably lower slope (compare the difference in abscissa scale with the two previous cases). The curve corresponding to $(\$_{ADC})_{\circ}(U_{P})$
also possesses a linear growth regime but with an even lower slope. And the curve corresponding to $(\$_{ADC})_{\circ}(U_{E})$  has completely abandoned the linear regime
and has the slowest entropy growth.
\par In summary the depolarizing channel (whose spectrum possesses the maximum degeneracy) is insensitive to the classical dynamics of the map it acts on. For the phase damping 
channel ( with a slightly less degenerate spectrum), the elliptic cat map ( with a regular classical behaviour) shows a slightly lower rate of entropy
production. For the amplitude damping channel (with its spectrum barely degenerate) all four entropy production rates are much slower than the two previous
cases; and there is also a strong dependence on the nature of the classical map,
with chaotic maps showing faster decoherence, which could be related to the idea of associating chaos to a quantum system according to how sensitive it is to
decoherence \cite{zurek-paz}.
\par As for the case of the superoperators acting on Grover's transformation, here it is also possible to recover from the spectra the conclusions just drawn above from
the entropies. In Fig. \ref{todoslosespectros} we can see the spectra of all four unitary maps in the (from up to down) first row and of $\$_{DC}$, $\$_{PDC}$  and $\$_{ADC}$
 composed with them in the second, third and fourth rows, respectively. The (from left to right) first column corresponds to the map $U_{HC}$, the second to $U_{PC}$, the
 third to $U_{EC}$ and the fourth to $U_{B}$. We can see how $\$_{DC}$ contracts the spectra radially leaving all eigenvalues but the $1$ on the circumference of radius
 $1-\epsilon$. $\$_{PDC}$ does approximately the same thing with most eigenvalues, but, on the other hand, for all four maps we find some eigenvalues that are left closer
 to the unitary circumference (which means that the behaviour of the composed map is closer to the unitary one's). Finally, we can see a strong enhancement of this tendency
 for the case of $\$_{ADC}$. When composed with the two hyperbolic maps we find more eigenvalues close to the unit circumference than in the two previous cases, but still
 concentrated at a distance $1-\epsilon$ from the origin. When composed with the parabolic map there are even more eigenvalues close to the unit circumference. And for the
 elliptic map the eigenvalues with modula close to unit are so many that they are seen in the plot as a continuum of eigenvalues accumulating on the unit circumference. So,
 also from this brief analysis, we can conclude that for the elliptic case the composed map remains the closest to the unitary one, behind it the parabolic map and last the
 two hyperbolic ones.
\par
\begin{figure}[ht!]
\begin{center} 
    \scalebox{.8}{%
    \includegraphics*[99,510][405,710]{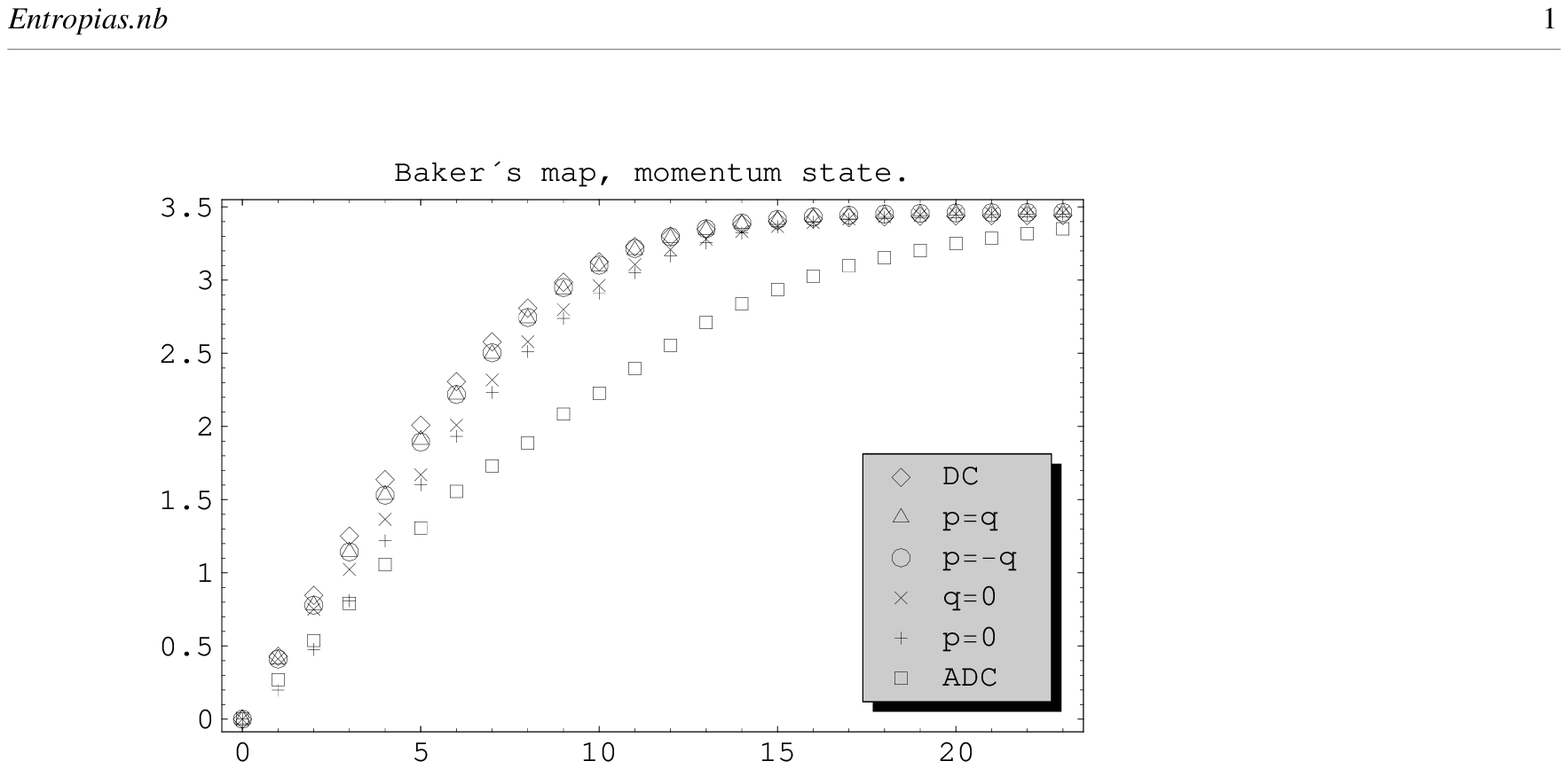}}
\end{center}
\begin{center} 
    \scalebox{.8}{%
    \includegraphics*[99,510][405,710]{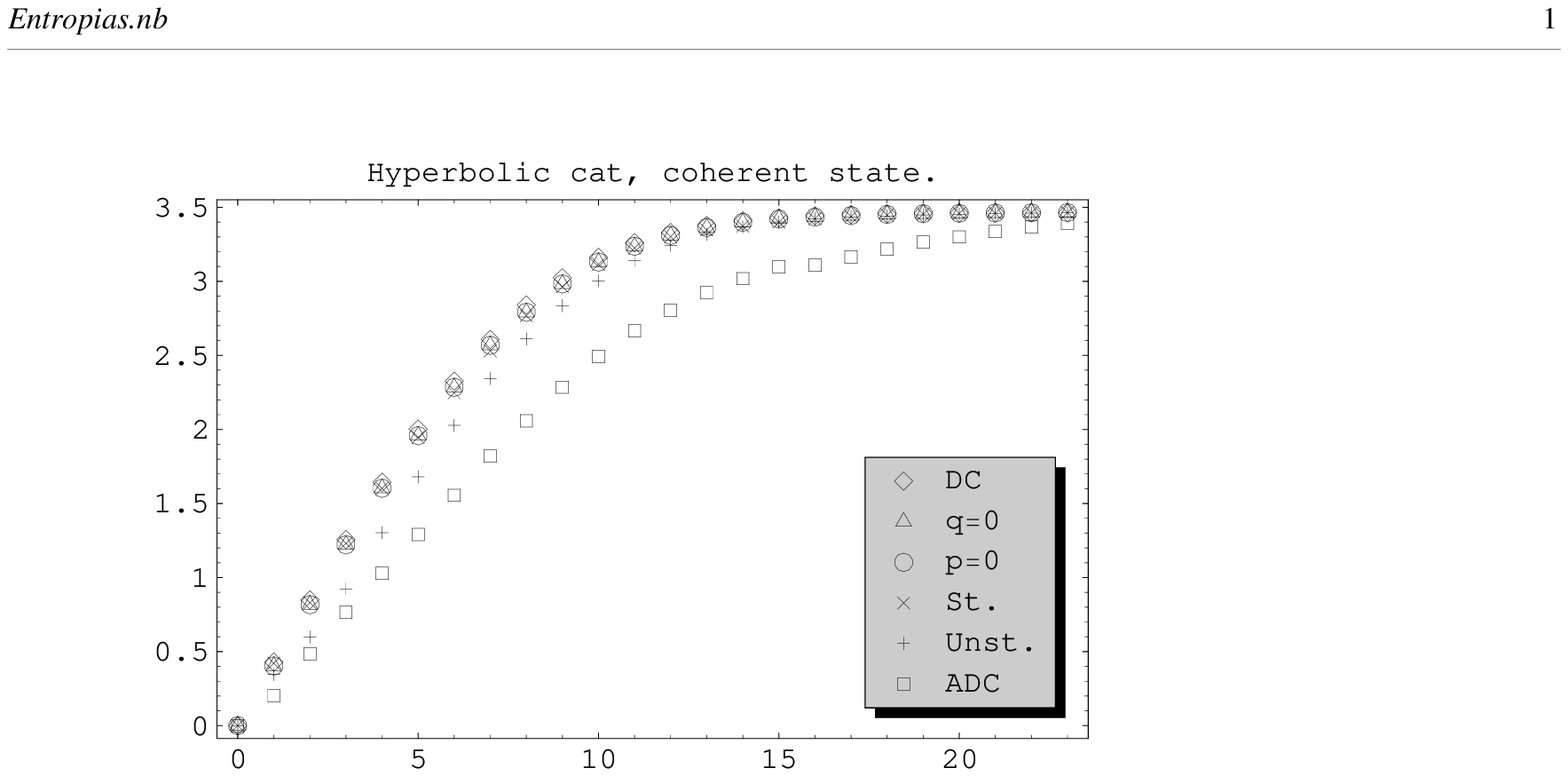}}
\end{center}
\begin{center} 
   \scalebox{.8}{%
    \includegraphics*[106,515][410,710]{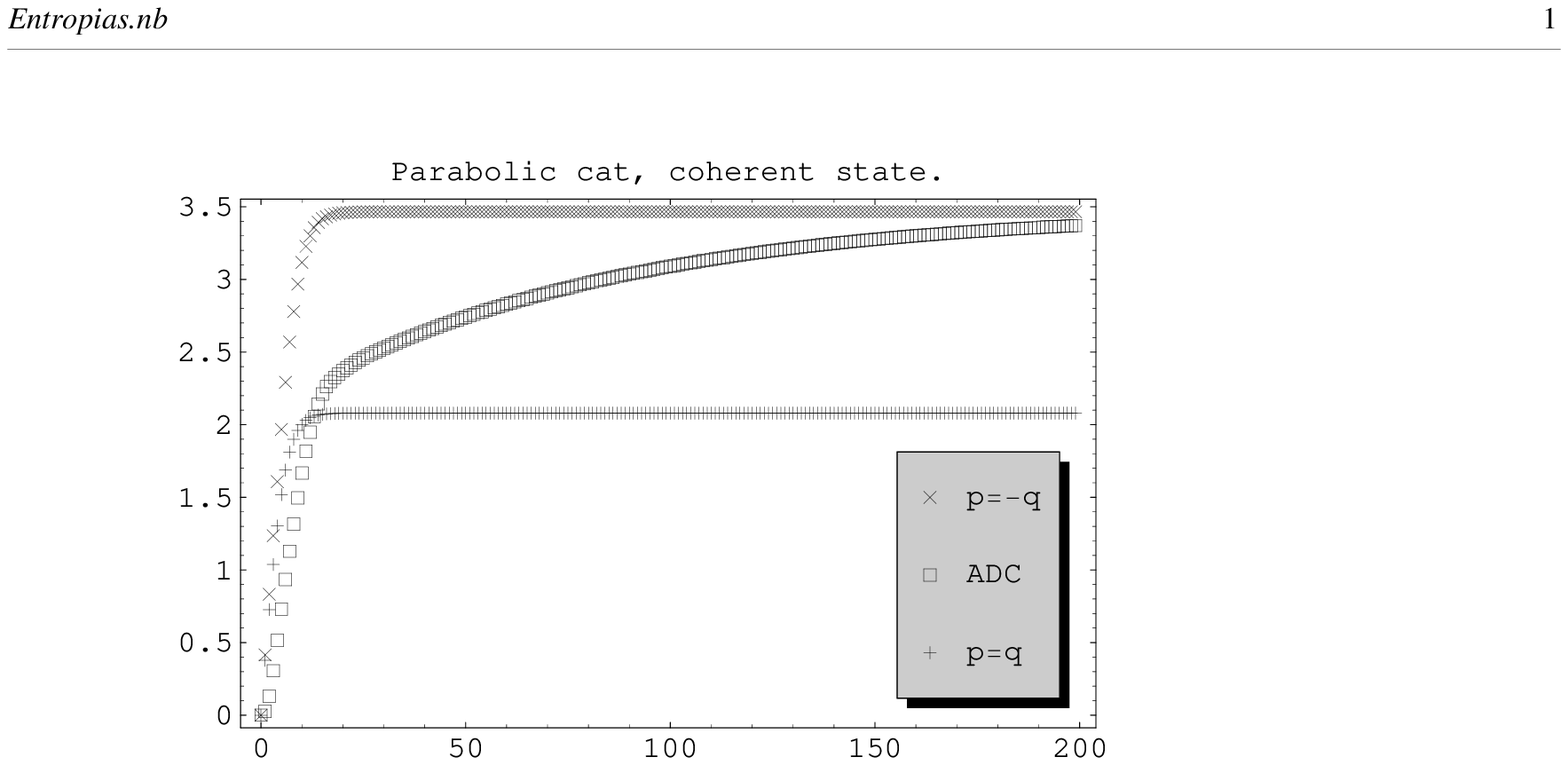}}
\end{center}
\caption{\footnotesize Linear entropy $S$ as a function of the number of iterations for the generalized depolarizing, phase damping (along several different lines) and
amplitude damping channels with the quantum baker's map (top), the quantum hyperbolic cat map (center) and the quantum parabolic cat map (bottom). In all cases the
parameters are $N=32$ and $\epsilon=0.2$. For the case of the baker's map the computer was initialized in the momentum state $\ket{p=0.25}$, while for the cases of the
hyperbolic and parabolic cat maps it was initialized in the coherent state $\ket{(q,p)=(0.25,0.25)}$. For the two fully chaotic cases there exists a slight dependence of
the entropy production rate on the diffusion direction.  This dependence is considerably stronger for the parabolic case, for which the number of iteration was taken
much greater so as  to fully appreciate the evolution of $S$ until it reaches saturation (note the difference in scales). (See text).\label{distdirofdif}}
\end{figure}

 Nevertheless, we have not yet exploited all the possibilities of the decoherent tools we have developed, for in the last section we have not appealed to the ability of
choosing the preferred basis of decoherence we possess through the generalized phase damping channel. For example, we can explore how the reaction of $U_{B}$ (and other
maps' too) depends on the direction of diffusion. And that is exactly what can be observed in Fig. \ref{distdirofdif}, where we have plotted the evolution of the entropy of
a system undergoing application of the composition of the depolarizing, phase damping (along several different lines) and amplitude damping channels with the quantum
baker's map (top), the quantum hyperbolic cat map (center) and the quantum parabolic cat map (bottom).
\par For the case of $U_{B}$ the system was initialized in a momentum state, the superoperators used were $\$_{DC}$, $\$_{L_{110}}$,
$\$_{L_{1-10}}$, $\$_{L_{010}}$, $\$_{L_{100}}$ and $\$_{ADC}$ . We can see that, as expected, the curve corresponding to $\$_{DC}$ is located on top of all others; and
that of $\$_{ADC}$ below all others. The novelty, though, is that in spite of all four different phase damping channels having the same degeneracy, there are two of them 
($\$_{L_{110}}$ and $\$_{L_{1-10}}$) whose curves locate right next to that of  $\$_{DC}$ and other two ($\$_{L_{010}}$ and $\$_{L_{100}}$) whose curves grow a little bit
more slowly. The only peculiarity in  $\$_{L_{100}}$ and $\$_{L_{010}}$ is that the directions of the diffusions they perform coincide with the classical map's invariant
directions (its stable and unstable manifolds). Equivalent results are obtained if the state is initialized in a position state. And also for the case of  $U_{HC}$, there
the initial state was taken to be a coherent one, and the superoperators used are the same except for the phase damping channels on oblique directions that have been
replaced by phase dampings on the lines of equations $100p=\pm 173q$, whose slopes are equal to $\pm 1.73\approx \pm \sqrt{3}$, the slopes of the classical map's stable
($+$) and
unstable ($-$) manifolds. Even though these two directions are just approximately equal to the map's invariant directions, we can observe in the case of the unstable
manifold a tendency for the entropy to remain below the others' (except for that of $\$_{ADC}$, of course).
\par There seems then to be a tendency of slower entropy growth when the diffusion is made along an invariant direction of the classical map. And we can see how this
tendency is considerably enhanced for the case of $U_{PC}$. There the initial state is also a coherent one and the superoperators taken are the same as those for the case
of $U_{B}$. Note how not only does the curve corresponding to the line of equation $p=q$ (the only invariant direction of the classical map) initially grows more slowly
than that of the line $p=-q$ (the curves corresponding to $\$_{DC}$ and to the lines of equation $q=0$ and $p=0$ have not been drawn for they are identical with the latter
one) but also how it deaccelerates so much that it even becomes lower than that of $\$_{ADC}$ and finally comes to a stop at a value much lower than the saturation value 
$S_{MAX}$ (note the difference in abscissa scale). So, in the same way as when we compared the maps' reactions before $\$_{ADC}$ (Fig. \ref{tresruid4maps}, bottom graph),
here for the parabolic case one also finds a much less decoherent reaction than that of the two fully chaotic cases. Then, also as in that case, one would expect this
tendency to be even more marked for the elliptic case; but we know that this map possesses no invariant directions. Moreover, for
this map the entropy was observed to be independent of the diffusion direction and all curves are identical to that observed in the central plot of Fig. \ref{tresruid4maps}
for the case of the simple phase damping channel.
\section{Conclusions}
   \label{Conclu}
   We have studied some superoperators that produce various kinds of non-unitary quantum operations that generalize the standard one qubit depolarizing, phase damping and amplitude
   damping channels to systems with arbitrary finite dimensions. The depolarizing and phase damping channels are diagonal in the chord basis consisting of
   phase space translation operators and can then be thought of as a special kind of
   random unitary process. This is not the case for the amplitude damping channel.  The efficiency by which these channels  take a
coherent superposition into different statistical mixtures was shown in phase space and was also understood from their spectra. The generalized amplitude damping channel
was found to possess a very weakly degenerate spectrum (in contrast to the completely degenerate spectrum of the depolarizing channel or the also very degenerate spectrum
of the phase damping channel).

\par As a first example of the action of these superoperators when combined with unitary maps 
we chose Grover's search algorithm. In the unitary case the algorithm possesses an invariant subspace and its operation is a simple rotation in this subspace. We find that our noise models do not alter 
substantially this picture. The invariant subspaces are now approximate as they are weakly coupled to the rest of the spectrum. As a result the search efficiency is altered both in a decrease 
of the success probability and in the value of the optimal iteration time. But,
nevertheless, the results indicate relatively good level of resistance to this models of noise, with success probability values ranging from $p_{e}\approx 0.25$ 
(for $\$_{DC}$) to $p_{e}\approx 0.5$ (for $\$_{PDC}$ and $\$_{ADC}$); for a decoherence parameter  $\epsilon=0.3$.
\par Another example of the effect of these superoperators was given by applying them to the quantum versions of classical maps possessing dynamics with different levels of
complexity: the baker's map and the hyperbolic, parabolic and elliptic cat maps. This time the reactions were analyzed from the point of view of their spectra and of the
linear entropy production rates induced by the composed maps.
The classical properties of the unitary maps are reflected in two ways in the composed spectrum: for hyperbolic maps,
level repulsion of the unitary map eigenvalues prevents degeneracies and therefore a given level of noise acts perturbatively producing a more or less uniform contraction of the spectrum towards a
modulus of order $1-\epsilon$. On the other hand integrable maps show in general many degenerate subspaces  that respond to the noise differently, the degeneracy being split in a non perturbative way and
leaving many eigenvalues close to the unit circle. In this case entropy is produced at a slower rate. For the phase damping on a line acting on hyperbolic and parabolic maps we find a dependence of the decoherence
rates with the alignment between the invariant manifolds and the line of diffusion.   
\par In more realistic models of decoherence acting on maps of mixed phase space the spectra may not be as simple and degenerate as the ones considered in this work, but still the analysis of the 
spectrum of the composed action will carry important information about the time behaviour of the open system.
 \begin{acknowledgments}
This work was financially  supported by CONICET and ANPCyT. M.L.A. wishes to thank the ``Pedro F. Mosoteguy''
Foundation for partial support.
\end{acknowledgments}


\begin{thebibliography}{99}
\bibitem{comentario}
From now on the words
``operator'' and ``matrix'' are to be used indistinctly.
\bibitem{chordrep}
An operator's chord representation is one in which it is expressed in terms of the discrete phase phace (DPS) translation operators, $\hat{T}_{(q,p)}$,
defined as $\hat{T}_{(q,p)}\ket{\mu}=e^{i\frac{2 \pi}{N}p(\mu+q/2)}\ket{\mu + q}$ and $\hat{T}_{(q,p)}\ket{\nu}=e^{-i\frac{2 \pi}{N}q(\nu+p/2)}\ket{\nu + p}$,
where $ \ket{\mu}$ and $\ket{\nu}$ stand for the $\mu^{th}$ and $\nu^{th}$ elements of the position and momentum bases, respectively. See
\cite{Miquel,ozorivas,schwinger}.
\bibitem{comentario2}
In fact, expanding $\hat{\rho}$ in the basis  $\{\hat{T}_{(q,p)}\}$ and using the fact that any $\hat{T}_{(q,p)}$ is
an eigenoperator of any diffusive superoperator (in particular of $\tilde{\$}_{DC}$) \cite{Yo}, it is easy to show that for any properly normalized density matrix      it is
$\hat{\rho}'\equiv\$_{DC}(\hat{\rho})\equiv (1-\epsilon)\hat{I}\hat{\rho}\hat{I}+\epsilon\tilde{\$}_{DC}(\hat{\rho})=(1-\epsilon)\hat{\rho}+\epsilon\frac{\hat{I}}{N}$.
\bibitem{comentario3}
A line $L_{n_{1},n_{2},n_{3}}$ in the DPS is a set of points $(q,p)$ in the  $N\times N$ grid $G_{N}$ defined such that $L_{n_{1},n_{2},n_{3}}=\{(q,p)\in
G_{N};n_{1}p=n_{2}q+n_{3},n_{1},n_{2},n_{3}\in\Bbb{N}\}$, where all sumations and equalities are to be understood as ``modulo $N$'' (See \cite{Miquel}).
\bibitem{comentario4}
Without  any loss of generality, any superoperator $\$$ can be expressed as 
$\$=\sum_{\alpha,\beta=0}^{N^{2}-1}s_{\alpha\beta}\hat{T}_{\alpha}\otimes\hat{T}^{\dagger}_{\beta}$, being $s_{\alpha\beta}$ a positive
definite $N^{2}\times N^{2}$ complex matrix \cite{caves}. But only those with a strongly degenerate spectrum (like $\$_{DC}$ and $\$_{PDC}$)
have a Kraus form free enough so as to remain diagonal when rewritten in another basis, for instance that of the $\hat{T}_{\alpha}$.
\end{thebibliography}
\end{document}